\documentclass[aps,pra,reprint,amsmath,amssymb,twocolumn]{revtex4} 
\usepackage{bbm}
\usepackage{graphicx}
\usepackage{bm}
\usepackage{amsmath}
\usepackage{amssymb,amsthm}
\usepackage{color}
\usepackage{hyperref}
\usepackage[sans]{dsfont}
\usepackage{placeins}
\newcommand{\ket}[1]{|#1\rangle}
\newcommand{\braket}[2]{\langle{#1}|{#2}\rangle}

\newcommand{\bra}[1]{\langle#1|}
\renewcommand{\S}{\mathcal{S}}
\def\eq{\begin{eqnarray}}
\def\en{\end{eqnarray}}
\usepackage[normalem]{ulem}
\newcommand{\perm}[0]{\text{perm}}

\begin{document}

\title{Sampling of partially distinguishable bosons and the relation to the multidimensional permanent}
\author{Malte C. Tichy}

\address{Department of Physics and Astronomy, Aarhus University, DK--8000 Aarhus, Denmark}

\begin{abstract}
The collective interference of  partially distinguishable bosons in multi-mode networks is studied via  double-sided Feynman diagrams. The probability for many-body scattering events  becomes a multi-dimensional tensor-permanent, which interpolates between distinguishable particles and identical bosons, and easily extends to  mixed initial states. 
   The permanent of the distinguishability matrix, composed of all mutual scalar products of the single-particle mode-functions, emerges as a natural measure for the degree of interference: It yields a bound on the difference between event probabilities for partially distinguishable bosons and the idealized species,    and exactly quantifies the degree of bosonic bunching. 
\end{abstract}
\date{\today}
\maketitle

\section{Introduction} 
Few physical problems fit in  the categories of computational complexity theory \cite{Moore:2011fk}. 
  An outstanding example for a fruitful interface between physics and computer science is Boson-Sampling \cite{Aaronson:2011kx}, the simulation of many indistinguishable bosons that scatter through a randomly chosen linear network with many more modes than particles: As a physical problem \cite{introBosonsSampling},  it is implemented straight-forwardly  with single photons \cite{Scheel:2004uq,introBosonsSampling,tichyTutorial}, as  demonstrated experimentally \cite{Spring15022013,Tillmann:2012ys,Crespi:2012vn,Broome15022013}. Mathematically, the probability for an output  event  equals the absolute square of the permanent of the  scattering sub-matrix, which has  well characterized complexity  \cite{Valiant:1979fk,Jerrum:2004:PAA:1008731.1008738,Aaronson:2011uq,Aaronson:2011kx,Aaronson:2013kx}. The  simplicity of the mathematical expression for the observable  physical quantity ensures that  Boson-Sampling remains simple enough to allow strong complexity-theoretic statements: It  is  very likely too complex to be solved approximately for any classical computer in polynomial time in the number of particles \cite{Aaronson:2011kx}. Whether a functional scalable Boson-Sampler  jeopardizes the extended Church Turing thesis \footnote{``Any efficient computation performed by a physical device can also be performed efficiently by a classical computer.''} is under debate \cite{willeverCT,ShchesnovichChurchTuring}, but consensus exists that Boson-Sampling constitutes a paradigm for a classically hard computational problem that is efficiently solved by a quantum physical system.

From a physical perspective, the successful experimental implementation of Boson-Sampling \cite{Spring15022013,Tillmann:2012ys,Crespi:2012vn,Broome15022013} has raised questions encompassing the scalability \cite{Motes:2013ys} and tolerance towards errors \cite{Leverrier:2013ys,Shchesnovich2013,PhysRevA.85.022332,ShchesnovichChurchTuring}, the generalization to experimentally more accessible systems \cite{Shen:2013zr,Lund:2013eb} and alternative physical implementations of the original problem \cite{PhysRevLett.113.120501,Rohde:2013vn,olson2014,sedhardreesan,Tamma2014}. Bridging the fields of physics and computer science, the problem of \emph{verifying} the functionality of an alleged Boson-Sampler arose \cite{gogolin2013,Aaronson:2013ls,aolitareliable}. While from a skeptical computer-science perspective, an efficient, loophole-free and unambiguous certification is  impossible \cite{gogolin2013,Aaronson:2013ls}, there are plausible  certification methods  based on physical properties such as bosonic statistics \cite{Carolan:2013mj}, probabilistic Bayesian reasoning \cite{Spagnolo:2013eu} and analytically solvable instances \cite{Tichy:2013lq,tichyTutorial}. 

The conditions assumed to establish the hardness of Boson-Sampling \cite{Aaronson:2011kx} -- low particle density, indistinguishable bosons, and random scattering matrix -- are crucial, since their violation enables efficient approximations: In the high-density limit of many more bosons than modes, semi-classical approaches become efficient \cite{JuanDiego,Shchesnovich:2013uq}. For distinguishable particles, one speaks of \emph{classical sampling},  which can be simulated inexpensively with a naive Monte-Carlo method that treats particles independently, one after another \cite{tichyTutorial,Jerrum:2004:PAA:1008731.1008738}. When artificial symmetries structure the scattering matrix, selection rules efficiently predict the strict suppression of certain output events \cite{tichyTutorial,Spagnolo:2013fk}. In these examples, physical intuition and the understanding of computational complexity nicely complement each other \cite{Gard:2013fk}.

Here, we pave the road to study the \emph{distinguishability transition} that connects ideal Boson-Sampling and its classical counterpart with distinguishable particles. On the one hand, the description of experiments requires a thorough understanding of such intermediate situation, because the indistinguishability of interfering bosons can be ensured only to a certain degree. On the other hand, numerous  fundamental open questions regarding partially distinguishable particles remain: How complex is the sampling problem on the \emph{transition} between the classical and the boson-sampler? Is there an unambiguous general quantifier of interference capability? Can one predict  how ``close'' a situation resembles the classical or the bosonic setup, and estimate the error due to neglecting or idealizing bosonic interference? 

Treatments of partially distinguishable particles formulated in the pure-state formalism \cite{Xiang:2006uq,Ou:2006ta,al:2007fk,Ou:2008zv,TichyFourPhotons,Ra:2013kx,younsikraNatComm,tichyTutorial,RohdeNew2014}  are not ideally suited to answer these questions, as  they suffer from computational costs and interpretational problems, exposed in Section \ref{partiallydist}. The many-particle scattering probability has more manageable expressions within the recently introduced density-matrix approach \cite{ShchesnovichPartial2014}, which naturally encompasses mixed initial states. Taking this method  \cite{ShchesnovichPartial2014} as a starting point, we focus on pure initial states in Section \ref{tensorperapproach}, present the -- to our knowledge --  most efficient algorithm for the computation of probabilities of pure partially distinguishable particles, and propose an interpretation based on double-sided Feynman diagrams. In Section \ref{degreeofdist}, a quantitative measure for the degree of interference is introduced, the permanent of the positive-definite hermitian \emph{distinguishability matrix}, which yields  bounds on the difference between the event probabilities for bosonic or distinguishable and partially distinguishable particles under certain assumptions. Furthermore, this measure exactly quantifies the enhancement of bunching events with all particles in one output port, and restricts the total variation distance between the respective probability distributions. The formalism allows us to explore and understand several counter-intuitive features of partially distinguishable particles: Perfect cancellation of amplitudes is not restricted to ideal bosons, but also occurs for partially distinguishable particles, as discussed in Section \ref{exampleperfectsupp}. Furthermore, simplistic intuitive Ans\"atze that interpolate between indistinguishable and distinguishable particles are ruled out in Section \ref{visualization}: Partially distinguishable situations do not resemble neither of the extreme cases, nor any mixture between them. 

\section{Partially distinguishable particles} \label{partiallydist}
Partially distinguishable particles have been treated by generalizing the bosonic permanent and the fermionic determinant to \emph{immanants} \cite{Tan:2013ix,Tillmann:2014ye,de-Guise:2014yf}. Alternatively, the initial many-body state can be decomposed into a sum of orthogonal terms with well-defined degrees of distinguishability, such that any two particles will either perfectly interfere or not at all  \cite{TichyFourPhotons,Ra:2013kx,younsikraNatComm,tichyTutorial,RohdeNew2014}. Recently, an approach based on the density-matrix formalism was proposed \cite{Shchesnovich2013,ShchesnovichPartial2014}, on which our calculations further below are based. 

The  tools used in Refs.~\cite{Tan:2013ix,Tillmann:2014ye,de-Guise:2014yf} to generalize the bosonic permanent to the immanant are admittedly a beautiful application of group theory and make the  symmetry breaking due to partial distinguishability tangible and illustrative. As pointed out in Ref.~\cite{ShchesnovichPartial2014}, however, these methods are not easily scalable, but force us to establish the transition probabilities anew for each particle number. As shown in Section \ref{orthonorsection} below, the orthonormalization of the single-particle mode functions \cite{TichyFourPhotons,Ra:2013kx,younsikraNatComm,tichyTutorial} reliably yields the desired many-particle transition probability in a scalable way,  but it comes with  high computational costs, interpretational issues, and without any straightforward generalization to mixed initial single-particle states. The complicated formulation aggravates any attempt to establish the actual computational complexity of the problem. 

\subsection{Scattering scenario}
\begin{figure}[th]
\includegraphics[width=\linewidth,angle=0]{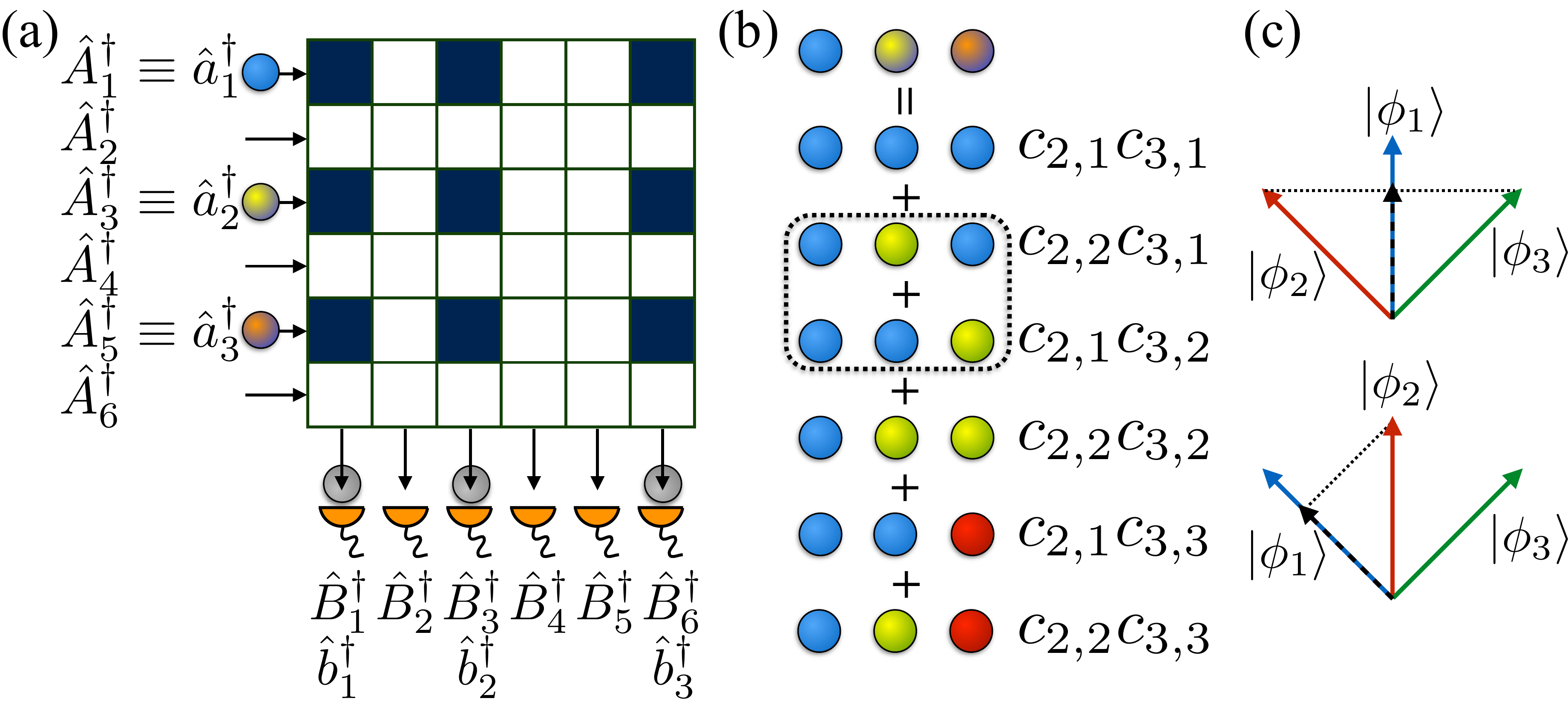}
\caption{(Color online) (a) Scattering matrix. Particles prepared in $\vec r=(1,0,1,0,1,0)$ scatter off the setup described by a matrix $U$; they are measured in the final configuration $\vec s=(1,0,1,0,0,1)$. The submatrix $M$ [Eq.~(\ref{matrixdef})] is highlighted in dark blue. For convenience of notation, we re-label the occupied modes from $1$ to $n$, depending on the initial and final state. Rows (columns) correspond to amplitudes related to input (output) modes. (b) Orthonormalization of a generic many-body state with three partially distinguishable particles, Eq.~(\ref{orthonormalizedfull}). The two components highlighted by dashed lines contribute to the same final state, interference between them needs to be taken into account. (c) Ambiguity of the fully indistinguishable weight $W_{\text{id}}=\braket{\tilde \phi_1}{\phi_2} \braket{\tilde \phi_1}{\phi_3}$, Eq.~(\ref{weightind}). Using the upper labeling, we find $W_{\text{id}}=1/4$; re-labeling the states as in the row below leads to $W_{\text{id}}=0$. In other words, the projection of all vectors onto $\ket{\phi_1}$ depends on the labeling of the vectors. The figure reproduces elements from \cite{tichyTutorial}. }  \label{setup.pdf}
 \end{figure}

Closely following the physical scenario exposed in Refs.~\cite{Tichy:2012NJP,tichyTutorial},  consider $n$ bosons prepared in the input modes of a scattering setup characterized by a unitary $m\times m$ matrix $U$. The initial distribution of the particles is defined by the \emph{mode occupation list}  $\vec r=(r_1, \dots, r_m)$ \cite{Tichy:2012NJP}, where $r_j$ particles populate input mode $j$. We are interested in the probability to find the final state $\vec s=(s_1, \dots, s_m)$ in the output modes, where $0 \le s_j , r_j \le n, \sum_{j} s_j = \sum_j r_j =n$. Since unoccupied input and output modes are irrelevant for the fate of the particles, we define the effective scattering   matrix $M$ as the relevant submatrix of $U$ that contains those rows and columns corresponding to initially and finally populated  modes, such that  the multiplicity of  rows and columns reflects the respective population. Using the \emph{mode assignment list} $\vec d(\vec s)=(d_1, \dots d_n)$ \cite{Tichy:2012NJP}, which indicates the mode in which the $j$th particle resides,  the effective scattering matrix becomes 
\eq
M= U_{\vec d(\vec r), \vec d(\vec s) } , \label{matrixdef}
 \en
 where our convention identifies  the $j$th row (column) with the $j$th input (output) mode, as illustrated in Fig.~\ref{setup.pdf}(a). 
 
Boson-Sampling \cite{Aaronson:2011kx,introBosonsSampling,tichyTutorial} constitutes a special instance of this general scattering problem: Each input mode is populated by at most one boson and $m \gg n$, which makes events with more than one particle per output mode  improbable. The unitary scattering matrix $U$ is picked randomly,  weighted by the Haar-measure. 
The resulting effective scattering matrix $M$ has no repeated rows and columns or any other structure, which impedes any simplification of the problem based on such possible multiplicity -- a mandatory pre-requisite for the argument that Boson-Sampling is computationally hard.

In the experiment, particles are typically not perfectly identical, but they carry degrees of freedom by which they can be distinguished to a certain extent  \cite{toeppel}. Distinguishability is described in the most general fashion by pooling \emph{all} possibly distinguishing degrees of freedom of the particles in the $j$th input mode into an ``internal'' state $\ket{\Phi_j}$ (particles in the same spatial mode are fully indistinguishable, which is fulfilled well in experiments with photons). In other words, $\ket{\Phi_j}$ contains all information that may potentially allow to distinguish the particle in mode $j$ from a particle in another mode $k$ other than the mode number itself, e.g.~frequency, polarization, time of arrival or spin. The mutual distinguishability of each of the $(n-1)n/2$ possible pairs out of $n$ particles is encoded in the hermitian positive-definite $n\times n$ \emph{distinguishability matrix}, 
\eq 
\mathcal{S}_{j,k} = \braket{\Phi_{d_j(\vec r)} }{\Phi_{d_k(\vec r) } } , \label{distinguishabilitymatrix}
\en
where $\mathcal{S}_{j,j}=1$. The multiplicity of rows and columns in the distinguishability matrix are equal and reflect the  multiple occupation of input modes.  For  $\mathcal{S}=\mathbbm{E}$ with $\mathbbm{E}_{j,k}=1$ for all $j$ and $k$, the problem reduces to  idealized boson-sampling, all particles are perfectly identical;  fully distinguishable particles, as they occur in classical sampling, are characterized by $\braket{\Phi_j}{\Phi_k}=\delta_{j,k}$, i.e.~$\S=\mathbbm{1}$. The approaches in Refs.~\cite{de-Guise:2014yf,ShchesnovichPartial2014,RohdeNew2014,tamm1,tamm2} also introduce matrices to quantify distinguishability (see Eq.~(C2) in \cite{de-Guise:2014yf}, Eq.~(6) in \cite{ShchesnovichPartial2014} and Eqs.~(20),(22) in \cite{RohdeNew2014}), the four definitions differ formally; for pure states, however, they contain  the same information and can be related to each other. 

Having established the physical setup, we formulate the scatting problem: The initial quantum state 
\eq 
\ket{\Psi_{\text{ini}}} &=& \prod_{j=1}^m \frac{1}{\sqrt{ r_j! }}  \left(  \hat A^\dagger_{j, \ket{\Phi_j} }  \right)^{r_j} \ket{0} \nonumber \\
&  =&  \frac{1}{\sqrt{ \prod_{k=1}^m r_k! }}    \prod_{j=1}^n    \hat A^\dagger_{d(\vec r)_j, \ket{\Phi_{d(\vec r)_j}} }   \ket{0} ,  \label{initialstate}
\en
 evolves via the single-particle transformation
\eq
\hat A^\dagger_{j, \ket{\Phi} } \rightarrow \hat U \hat A^\dagger_{j, \ket{\Phi} }   \hat U^{-1} = \sum_{k=1}^m U_{j,k} \hat B^\dagger_{k, \ket{\Phi}}  ,
\en
where $\hat U$ describes the unitary evolution induced by the multimode setup \cite{tichyTutorial} and $\hat B^\dagger_{k,\ket{\Phi}}$ creates a particle in   output mode $k$ and in  the internal state $\ket{\Phi}$. 
The final state becomes 
\eq 
\ket{\Psi_{\text{fin}}}= \hat U \ket{\Psi_{\text{ini}}} .
\en
Our object of interest is the probability $\mathcal{P}_{\mathcal{S}}( \vec s)$ to find the output configuration $\vec s$ for particles characterized by  $\mathcal{S}$. Our notation omits the initial configuration $\vec r$, which is assumed to be fixed. 
 For convenience, occupied input and output modes are relabeled by $1, \dots, n$, i.e.~$\hat a^\dagger_{k} \equiv \hat A^\dagger_{d_k(\vec r)}, \hat b^\dagger_{k} \equiv \hat B^\dagger_{d_k(\vec s)}, \ket{\phi_k} \equiv \ket{\Phi_{d_k(\vec r)}}$, which amounts to formally assuming $ \vec r=\vec s=(1,1, \dots, 1, 0, \dots ,0)$ [Fig.~\ref{setup.pdf}(a)]. This convention makes all appearing expressions easier to read.

\subsection{Treatment via selected single-particle basis} \label{orthonorsection}
In the following, we argue that no single-particle basis allows a satisfactory treatment of the problem in the pure-state formalism: 
Given a single-particle basis $\{ \ket{ \eta_1}, \dots, \ket{\eta_n} \}$ that spans the generally $n$-dimensional internal Hilbert-space containing $\{ \ket{\phi_1}, \dots, \ket{\phi_n} \}$, each $\ket{\phi_k}$ in Eq.~(\ref{initialstate}) is a superposition of all $n$ $\ket{\eta_j}$, which makes the many-body state (\ref{initialstate}) a sum of $n^n$ orthogonal terms. The number of non-vanishing terms in this expansion is reduced to $n!$ by selecting the basis $\{ \ket{\tilde \phi_1} , \dots , \ket{\tilde \phi_n} \}$ obtained  via Gram-Schmidt-orthonormalization of $ \{ \ket{\phi_1} , \dots , \ket{\phi_n} \} $ \cite{TichyFourPhotons,Ra:2013kx,younsikraNatComm,tichyTutorial}, which ensures that $k>l \Rightarrow \braket{\tilde \phi_k}{\phi_l} =0 $. In particular, if the $n$ particles only span a $D<n$-dimensional Hilbert space,  at most $D^{(n-D)} D! $ orthogonal terms remain. Formally, by writing the single-particle states as linear combinations of the $\ket{\tilde \phi_k}$, 
\eq 
\ket{\phi_j} = \sum_{k=1}^j c_{j,k} \ket{\tilde \phi_{k}} ,
\en
we express  the full many-body state (\ref{initialstate}) as  \cite{tichyTutorial} \eq 
\ket{\Psi_{\text{ini}}}= \sum_{k_2=1}^2 \dots  \sum_{k_n=1}^n \hat a^\dagger_{1, \ket{\tilde \phi_{1}}} \left( \prod_{j=1}^n c_{j,k_j}    \hat a^\dagger_{j, \ket{\tilde \phi_{k_j}} }  \right)  \ket{0} .  \label{orthonormalizedfull}
\en
The orthonormality of the $\ket{\tilde \phi_j}$ then allows us to write the probability for the event $\vec s$ as \cite{TichyDiss} 
\eq 
\mathcal{P}_{\mathcal{S}}(\vec s) = \nonumber \hspace{6.7cm} \\ 
 \sum_{j_2=1}^2 \sum_{j_3=1}^3 \dots \sum_{j_n=1}^n  
\sum_{\sigma \in S_{\{1, j_2, \dots , j_n \}} } \left|  \bra{ \Psi_{\text{fin}}(\vec j, \sigma) }  \hat U \ket{\Psi_{\text{ini}} }  \right|^2 , \label{orthoproba}
\en
where the sums over the $j_k$ take into account all configurations of particles in orthonormalized internal states that are found in the output ports and
\eq 
\ket{ \Psi_{\text{fin}}(\vec j, \sigma)  }  = \mathcal{\tilde N}(\vec r, \sigma) \prod_{k=1}^n    \hat b^\dagger_{k, \ket{\tilde \phi_{\sigma_k} } }   \ket{0}  ,
\en
is the final state with the particles in the internal states $\{ \ket{\tilde \phi_{1} }, \ket{\tilde \phi_{j_2} },  \dots , \ket{\tilde \phi_{j_n} } \}$ distributed among the output ports according to the permutation $\sigma$, where  $\mathcal{\tilde N}(\vec r, \sigma)$ accounts for the overnormalization due to output modes  occupied by several indistinguishable particles. 
In evaluating (\ref{orthoproba}), the terms in the initial and final states that do not contain the  same set of single-particle states $\{ \ket{\tilde \phi_{1}},  \ket{\tilde \phi_{2}}, \dots, \ket{\tilde \phi_{j_n}} \}$ vanish. A generalization of Eq.~(\ref{orthoproba}) to any basis $\{ \ket{\eta_1}, \dots, \ket{\eta_n} \}$ was recently presented in \cite{RohdeNew2014}.

Although the evaluation based on (\ref{orthoproba}) can be scaled to reasonably large particle numbers \cite{Tichy:2013lq} and allows a visual interpretation in terms of Feynman diagrams \cite{tichyTutorial}, there are four main  caveats that make it unsatisfactory: In the first place, no clear physical interpretation of the weights $c_{j,k}$ is possible, as these depend on the choice of the single-particle basis $\{ \ket{\tilde \phi_1}, \dots , \ket{\tilde \phi_n} \} $. In particular, it is tempting to interpret the amplitude of the fully indistinguishable component in Eq.~(\ref{orthonormalizedfull}), 
\eq 
W_{\text{id}}=\left| \prod_{j=2}^n \braket{\tilde \phi_1}{\phi_j} \right|^2= \prod_{j=2}^n |c_{j,1}|^2  \label{weightind} ,
\en 
as the ``perfectly interfering'' part of the wavefunction: It corresponds to the component of the wavefunction that features perfect many-body interference in Eq.~(\ref{orthoproba}), which allows one to formulate bounds to the deviation from the ideal bosonic probability distribution \cite{Tichy:2013lq}. However,  the set $\{ \ket{ \tilde \phi_1}, \dots , \ket{ \tilde \phi_n} \}$ depends on the ordering of vectors on which the orthonormalization is performed and the 
 weight (\ref{weightind}) varies considerably with the respective choice of basis. As an extreme example, sketched in Fig.~\ref{setup.pdf}(c), consider 
\eq \ket{\phi_1}=\ket{0}, \ket{\phi_2}=\frac{ \ket{0} + \ket{1} }{\sqrt 2}, \ket{\phi_3}=\frac{\ket{0} - \ket{1} }{\sqrt 2}  ,
\en
where $\ket{0}, \ket{1}$ denote any qubit-like degree of freedom. The orthonormalization yields 
\eq 
\ket{\tilde \phi_1} =  \ket{0}, \ \ket{\tilde \phi_2} = \ket{1}, 
\en 
for which the weight of the fully indistinguishable component becomes $W_{\text{id}}=|\braket{\tilde \phi_1}{\phi_2}\braket{\tilde \phi_1}{\phi_3} |^2 =1/4$. Exchanging the labels of the first and second ports does manifestly not modify the actual physical situation, 
\eq \ket{\phi_1}=\frac{ \ket{0} + \ket{1} }{\sqrt 2},  \ket{\phi_2}=\ket{0}, \ket{\phi_3}=\frac{  \ket{0} - \ket{1}  }{\sqrt 2} ,
\en
but leads to 
\eq 
\ket{\tilde \phi_1} = \frac{ \ket{0}+\ket{1} }{\sqrt 2}, \ \ket{\tilde \phi_2} = \frac{  \ket{0}-\ket{1} }{\sqrt 2}, 
\en 
and, consequently, $W_{\text{id}}=0$. In other words, a set of weights $c_{j,k}$ does not allow any immediate quantitative statement on the degree of interference in the system. 

As a second point, although all summands that contribute to the initial state in Eq.~(\ref{orthonormalizedfull}) are orthogonal, several feed the  same final state: For $n=3$, the components with weights $c_{2,1}c_{3,2}$ and $c_{2,2}c_{3,1}$ contain the same set of single-particle states $\{ \ket{\phi_1}, \ket{\phi_1}, \ket{\tilde \phi_2} \}$, and both contribute  to the same final states [see Fig.~\ref{setup.pdf}(b)] \cite{tichyTutorial}. As a consequence, interference between these two terms occurs and event probabilities depend on the relative phase between $c_{2,1}c_{3,2}$ and $c_{2,2}c_{3,1}$. The initial Fock-state (\ref{initialstate}), however, is free of any  phase relationship between the input modes, which makes the emerging relative phase a formal artifact, which further complicates the interpretation of the actual physical process. Summarizing these two points, the coefficients $c_{j,k}$ do not bear clear  physical meaning, they emerge as formal but unavoidable intermediate step between the physically meaningful scalar products and the observable event probabilities. Due to the orthonormalization, the dependence of the $c_{j,k}$ on scalar products is  intricate and obfuscates the behavior of event probabilities. 
  
Thirdly, the computational expenses required to treat Eq.~(\ref{orthoproba}) are considerable: Not only does each scalar product $\bra{\Psi_{\text{fin}}(\vec j, \sigma) } \hat U \ket{\Psi_{\text{ini}}}$ involve at least one permanent, but we also need to consider the superposition of terms in $\ket{\Psi_{\text{ini}}}$ that lead to the same final state. The total number of separable final states in the sum (\ref{orthoproba}) is $n!$; since each of the $n!$ appearing permanents requires an exponentially large number of operations, the total expenses scale dramatically. A precise quantification of the computational expenses for an approximate evaluation is difficult, as the number of terms depends on the chosen basis. 

As a fourth and last point, the extension to mixed states is not  straightforward: For 
  a mixture of internal states at input mode $j$, $\hat \varrho_j = \sum_{k} p_{j,k} \ket{\psi_{j,k}}\bra{\psi_{j,k}} $,  the orthonormalization needs to be performed for each set of pure states $\{ \ket{\psi_{1,k_1}},  \ket{\psi_{2,k_2}}, \dots , \ket{\psi_{n,k_n}}  \}$. For mixed states with high rank, the emerging  Hilbert space can have dimension much larger than $n$. 

Given these caveats, the question naturally arises whether there exists an optimal basis $\{ \ket{\eta_1} , \dots , \ket{\eta_n} \}$ that remains free of the above problems. For example, for $n=3$, we impose that there be only 5 orthogonal terms in the decomposition (\ref{orthonormalizedfull}) into single-particle states $\ket{\eta_j}$. As one  quickly realizes by imposing these requirements as boundary conditions, there is, in general, no basis with these desirable properties for  $n \ge 3$. In other words, any pure-state approach based on the sum of overlaps between possible final states with the time-propagated initial state of the form Eq.~(\ref{orthoproba}) suffers from the four caveats  mentioned  above. The  Gram-Schmidt orthonormalized basis $\{ \ket{\tilde \phi_1} , \dots , \ket{\tilde \phi_n} \}$ seems to be the best, yet unsatisfactory, choice for a single-particle basis. 

A more convenient representation of the event probability (\ref{orthoproba}) is desirable. In particular, the intermediate step via the coefficients $c_{j,k}$ is cumbersome and formal, and does not offer good physical insight. In the following,  an approach based on the density-matrix formalism \cite{ShchesnovichPartial2014}, even if it seemingly presents a complication of the problem at first sight, leads to a compact form for the transition probabilities and  naturally solves the exposed problems.

\section{Tensor-permanent approach} \label{tensorperapproach}
\subsection{Event probability as expectation value} \label{evprobexp}
The discussion in the last section motivates us to seek a representation of event probabilities as a function of the scalar products of the single-particle mode functions, i.e.~as a function of the matrix elements of the distinguishability matrix $\S$ defined in Eq.~(\ref{distinguishabilitymatrix}). Such representation emerges by expressing the measurement of the particle arrangement $\vec s$ by the high-dimensional operator that projects onto the space with one particle per output mode \emph{without} differentiating the internal states \cite{ShchesnovichPartial2014}, 
\eq 
\hat P_{\mathbbm{1}} & =&  \sum_{X_1, \dots , X_S } \prod_{j=1}^n \hat b_{j, \ket{X_j} }^\dagger \ket{0} \bra{0} \prod_{j=1}^n \hat b_{j, \ket{X_j}}  , \label{projector}
\en
where the sum over the $X_j$ runs over all states of a basis that span the ``internal'' Hilbert space, i.e.~for all $k$, $\sum_j  \braket{\phi_k} {X_j} \braket{X_j}{\phi_k} =1$. For continuous degrees of freedom, the sum needs to be replaced by an integral over the respective basis states. The computation of event probabilities in Ref.~\cite{Spagnolo:2013fk} was implicitly based on this approach, and a similar projection operator is also used in Ref.~\cite{de-Guise:2014yf}. The event probability $\mathcal{P}_{\S}(\vec s)$ is  the expectation value of this  projector, 
\eq 
\mathcal{P}_{S}(\vec s) &=& \mathcal{N}~ \bra{\Psi_{\text{fin}}} \hat P_{\mathbbm{1}} \ket{\Psi_{\text{fin}}}   \label{transprob} ,
\en
where the normalization factor $\mathcal{N} = 1 /( {\prod_j s_j! r_j! }) $ becomes necessary due to our incorporation of multiply occupied input and output modes via the multiplicities of the respective rows and columns in the scattering matrix (\ref{matrixdef}). For convenience of notation, the normalization factor is omitted in the following by assuming -- unless explicitly mentioned otherwise -- that the initial and final states do not contain multiply populated modes.

The projection of the final wavefunction in the eigenspace of $\hat P_{\mathbbm{1}}$ is the component with precisely one particle per occupied output mode. It becomes a superposition of the particles among the modes, 
\eq
\nonumber 
\ket{\Psi_{\text{coinc}}} &=& \sum_{\sigma \in S_n} \prod_{j=1}^n M_{\sigma_j, j} \hat b_{j, \ket{ \phi_{\sigma_j} } }^\dagger \ket{0}  \nonumber \\
 & = & \hat P_{\mathbbm{1}} \hat U \ket{\Psi_{\text{ini}}}  = \hat P_{\mathbbm{1}} \ket{\Psi_{\text{fin}}} , \label{finalstate}
\en
i.e.~the particle that was originally prepared in mode $\sigma_j$ ends in mode $j$, and carries its internal degree of freedom $\ket{\phi_{\sigma_j}}$.  The state (\ref{finalstate}) is (besides the trivial case $M=\mathbbm{1}$) sub-normalized, its norm yields the desired probability to find the  distribution of particles $\vec s$ in the output modes:
\eq
\mathcal{P}_{\S}(\vec s) &=& 
 \bra{\Psi_{\text{ini}}} \hat U^\dagger \hat P_{\mathbbm{1}} \hat U \ket{\Psi_{\text{ini}}}  \label{propaandback}  \\
&=& \nonumber  
\braket{\Psi_{\text{coinc}}}{\Psi_{\text{coinc}}}  \\ 
&=&   
 \sum_{\sigma, \rho \in S_n}  \prod_{j=1}^n  \left(  M_{\sigma_j, j} M^*_{\rho_j, j}  \mathcal{S}_{\rho_j,\sigma_j}  \right)  \label{bestphysinter} 
\en
By defining the $n^3$-dimensional 3-tensor
\eq 
W_{k,l,j}=M_{k,j} M^*_{l,j} \mathcal{S}_{l,k} , \label{Wtensor}
\en
 the event probability becomes a multi-dimensional tensor permanent \cite{Barvinok2014,Potapov2011}, 
\eq 
\mathcal{P}_{\S}(\vec s) &=& 
 \text{perm}(W) = 
 \sum_{\sigma, \rho \in S_n} \prod_{j=1}^n W_{\sigma_j, \rho_j, j}  , \label{bigper}
\en
which generalizes  the permanent of a matrix (2-tensor).  In the representations (\ref{bestphysinter}) and (\ref{bigper}), the roles of $M$, $M^*$ and $\S$ are formally equivalent; these three matrices can be permuted at will to yield different equivalent expressions: 
\eq
\mathcal{P}_{\S}(\vec s)  &= & 
 \sum_{\rho \in S_n} \left[ \prod_{j=1}^n \mathcal{S}_{j,\rho_j }  \right] \text{perm}\left( M* M^*_{\rho,\mathbbm{1}}  \right) \label{eq:fullexp} \\
&= & 
\sum_{\rho \in S_n} \left[ \prod_{j=1}^n M_{\rho_j,j}  \right] \text{perm}\left( \mathcal{S}^{*}_{\mathbbm{1},\rho} * M^*  \right) \label{eq:fullexp2}  \\
&= & 
 \sum_{\rho \in S_n} \left[ \prod_{j=1}^n M^*_{\rho_j,j}  \right] \text{perm}\left( \mathcal{S}_{\mathbbm{1},\rho} * M  \right) \label{eq:fullexp3}  ,
\en
where $A_{\rho,\sigma}^*$ denotes the complex-conjugate of the matrix $A$ with rows permuted according to $\rho$ and columns permuted according to $\sigma$, and $*$ denotes the entrywise Hadamard-product. 

\subsection{Distinguishable particles and identical bosons} \label{distpaidbos}
Identical particles  interfere  perfectly, such that the sums over $\sigma$ and $\rho$ in Eq.~(\ref{bestphysinter}) loose their mutual dependence, and 
 \eq
 \mathcal{P}_{\text{id}}(\vec s) \equiv  \mathcal{P}_{\mathbbm{E}}(\vec s) &=& \left( \sum_{\sigma \in S_n} \prod_{j=1}^n  M_{\sigma_j, j} \right)  \left(\sum_{\rho \in S_n} \prod_{j=1}^n    M^*_{\rho_j, j} \right) \nonumber  \\
 &=& \left| \sum_{\sigma \in S_n} \prod_{j=1}^n  M_{\sigma_j, j} \right|^2 = |\perm(M)|^2,  \label{idprob}
 \en
where we recover the probability for identical bosons. 
 The simplification of (\ref{idprob}) with respect to (\ref{bigper}) arises  due to the uniqueness of the final state  for indistinguishable particles, i.e.~the projector (\ref{projector}) has exactly one eigenvector. Using Eq.~(\ref{eq:fullexp}), 
\eq 
| \text{perm}(M) |^2 = \sum_{\sigma \in S_n} \text{perm}\left( M * M^*_{\sigma, \mathbbm{1}}  \right) ,
\en
which expresses the 1:1-relationship between single- and double-sided Feynman diagrams for coherent propagation. 

For fully distinguishable particles, $\S_{j,k}=\delta_{j,k}$, one sum over all permutations in Eq.~(\ref{bestphysinter}) collapses, and 
\eq
\mathcal{P}_{\text{dist}}(\vec s)\equiv \mathcal{P}_{\mathbbm{1}}(\vec s)& =& \sum_{\sigma \in S_n} \prod_{j=1}^n  \left( M_{\sigma_j, j} M^*_{\sigma_j, j}  \right) \nonumber \\
&= &\text{perm}( |M|^2 ), \label{distprobfull}
 \en
 i.e.~the permanent of the absolute-squared-matrix ${|M|^2 \equiv M* M^*}$, which 
  can be approximated efficiently thanks to the positivity of the matrix elements of $M* M^*$ \cite{Jerrum:2004:PAA:1008731.1008738}.

\subsection{Double-sided Feynman diagrams}
\begin{figure}[h]
\includegraphics[width=\linewidth,angle=0]{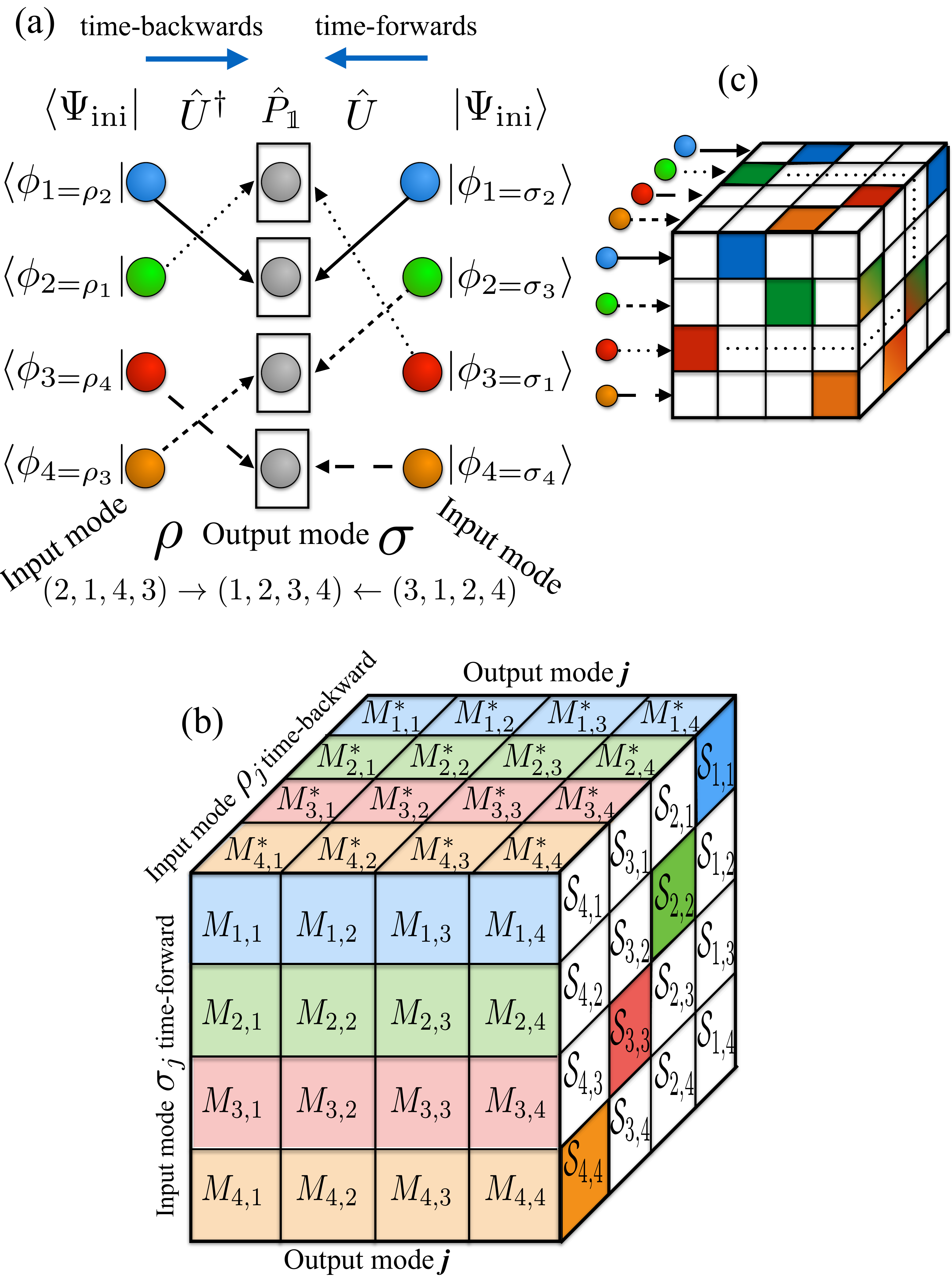}
\caption{(Color online) (a) Double-sided Feynman diagram corresponding to the product $M_{1,2} M_{2,3} M_{3,1} M_{4,4} M^*_{1,2} M^*_{2,1} M^*_{3,4} M^*_{4,3} \mathcal{S}_{1,1}  \mathcal{S}_{4,2}  \mathcal{S}_{2,3}  \mathcal{S}_{3,4}$. (b) Visualization of the tensor $W_{k,l,j}=M_{k,j} M^*_{l,j} \mathcal{S}_{l,k}$, defined in (\ref{Wtensor}).  Each one of the $n^3$ elementary cubes within the three-dimensional cube corresponds to one amplitude of a particle starting in a certain input mode, being detected in an output mode and traveling back in time to a possibly different input mode. (c) Position of the tensor elements corresponding to the Feynman diagram (a), one single-particle path is emphasized by the dotted line. 
 }  \label{FullVisualization.pdf}
 \end{figure}

Having established compact representations of the event probability $\mathcal{P}_{\S}(\vec s)$, we interpret these sums physically and visually \cite{diffediagnosis}. Eq.~(\ref{propaandback}) describes propagation forwards in time via $\hat U$, the subsequent projection on the desired subspace via $\hat P_{\mathbbm{1}}$, and propagation backwards in time via $\hat U^\dagger$. The sum over $\sigma$ in Eq.~(\ref{bestphysinter}) represents all possible paths that the particles take forwards in time [see Fig.~\ref{FullVisualization.pdf}(a)], $\rho$ then describes the paths backwards in time. Consequently, a particle starting in input port $\sigma_j$ is detected in output port $j$, to then propagate back into the initial port $\rho_j$. The amplitudes of the processes are $M_{\sigma_j, j}$ and $M_{\rho_j,j}^*$, respectively, which need to be amended further by  the overlap of the internal states $\braket{\phi_{\rho_j}}{\phi_{\sigma_j}}=\S_{\rho_j, \sigma_j}$, which explains the product of these three quantities in Eq.~(\ref{bestphysinter}). For $\sigma=\rho$, no bosonic exchange processes occur, which is characteristic for the classical, distinguishable contribution. All the possible processes need to be added; since the permutations $\sigma$ and $\rho$ describing time-forward and time-backward propagation are independent, we remain with a double-sum over $\sigma$ and $\rho$. 

The tensor $W_{k,l,j}$ is illustrated in Fig.~\ref{FullVisualization.pdf}(b): The element $W_{k,l,j}$ contains the amplitude for a particle ``starting'' in $k=\sigma_j$, being detected in $j$ and ``ending'' in $l=\rho_j$ after its time-reversed travel, each tensor element is therefore the product of the respective matrix elements of the single-particle unitary time evolution, attenuated by the scalar product of the internal state in the ``initial'' and ``final'' input mode. Each double-sided Feynman diagram corresponds to one \emph{diagonal} of the 3-tensor, i.e.~$n$ cubic elements that lie on different columns for each of the three dimensions. The highlighted elements in Fig.~\ref{FullVisualization.pdf}(c) are precisely those corresponding to the Feynman diagram in (a). 

Given this interpretation as many-particle paths, it is instructive to write the probability for partially distinguishable particles as a classical term for distinguishable particles (all paths for which the very same particles traveling forward and backward in time meet in the same output mode $j$, i.e.~$\rho=\sigma$ in Eq.~(\ref{bestphysinter})), which is attenuated by exchange processes, i.e.~non-classical many-particle paths, 
\eq 
\mathcal{P}_{\S}(\vec s)= \mathcal{P}_{\text{dist}}(\vec s) + \sum_{\rho \neq \mathbbm{1} }  \left[ \prod_{j=1}^n \mathcal{S}_{j,\rho_j }  \right] \text{perm}\left( M* M^*_{\rho,\mathbbm{1}}  \right) , \label{asnonclass}
\en 
where the number of fixed points of $\rho$, $|\{ k | k = \rho_k \}|$, counts the particles that do not participate in any exchange process. Interfering terms are bounded in magnitude by the classical contribution, $  \mathcal{P}_{\text{dist}}(\vec s) \ge |\text{perm}\left( M* M^*_{\rho,\mathbbm{1}}  \right)|$, as shown in Appendix \ref{maxiperm}. 

\subsection{Evaluation via Ryser's algorithm}
At first sight, Eq.~(\ref{bigper}) appears not particularly benevolent, since, instead of a sum over all $n!$ permutations as in the matrix permanent, it contains \emph{two} sums over $n!$ entries. The computational expenses for the transition probabilities are alleviated by closely following Ryser's algorithm \cite{Ryser:1963oa}. Applying the inclusion-exclusion principle, we find 
 \eq
\mathcal{P}_{\S}(\vec s)=  \sum_{\substack{S, R \subseteq \\ \{1, \dots , n\} } } (-1)^{|S|+|R|} \prod_{j=1}^n \sum_{\substack{ r \in R \\ s \in S}}M_{s,j} M^*_{r,j} \mathcal{S}_{r,s}  ,
 \en 
 i.e.~a sum over $2^{2n}$ terms instead of $n!^2$ as in Eq.~(\ref{bestphysinter}). Exploiting  $W_{j,k,l}=W_{k,j,l}^*$, we  eliminate approximately half of the terms, 
 \eq
\mathcal{P}_{\S}(\vec s)=  
 \sum_{S \ge R \subseteq \{1, \dots , n\} } (2-\delta_{S,R}) (-1)^{|S|+|R|}  \nonumber \times \\ 
  \Re\left\{  \prod_{j=1}^n \sum_{\substack{ r \in R \\ s \in S} } M_{s,j} M^*_{r,j} \mathcal{S}_{r,s}  \right\} , \label{ryserbest}
\en
where $S \ge R$ orders the subsets to avoid an evaluation of  both $(S,R)$ and $(R,S)$. Although Eq.~(\ref{ryserbest}) still bears considerable computational expenses, it reduces the computational costs to a level that will allow us to numerically explore the realm of a moderate number of partially distinguishable particles in Section \ref{visualization}, which remained unfeasible using Eq.~(\ref{bestphysinter}). 
 
\subsection{Perfect suppression for partially distinguishable particles} \label{exampleperfectsupp}
The formalism established in Section \ref{evprobexp} allows us to explore the realm of partially distinguishable bosons and understand its peculiarities. For example, it seems tempting and rather intuitive to assume that any event $\vec s$ with finite classical probability for distinguishable particles be also realized with finite probability whenever interference is not perfect, i.e.~
\eq 
\S \neq \mathbbm{E}, \mathcal{P}_{\text{dist}}(\vec s)  \neq 0  \Rightarrow \mathcal{P}_{\S}(\vec s) \neq 0  , \label{conjecture}
\en
which formalizes the intuitive idea that fully destructive interference only arises for perfectly indistinguishable particles. Alternatively, by defining the visibility of events that are fully suppressed for identical particles by destructive interference ($P_{\text{id}}(\vec s)=0$) using the probability for distinguishable particles as a point of reference,
\eq 
V = \left| \frac{\mathcal{P}_{\text{dist}}(\vec s) - \mathcal{P}_{\S}(\vec s)  }{\mathcal{P}_{\text{dist}}(\vec s) + \mathcal{P}_{\S}(\vec s)  }  \right|,
\en
 we are tempted to state that 
\emph{partial distinguishability implies imperfect visibility}, $\mathcal{S}\neq \mathbbm{E} \Rightarrow V \neq 1$. This is indeed true for two-photon Hong-Ou-Mandel interference   \cite{Hong:1987mz}. A naive extrapolation of single-particle wave-particle duality \cite{Englert:1996ys} to the many-body domain also seems to make such relationship plausible: Distinguishing information on the path taken by a particle jeopardizes wave-like interference visibility. 

A counter-example against conjecture (\ref{conjecture}) forces us to be careful with promoting our natural intuition based on single-particle interference to the many-particle realm: Into a setup with $m=9$ modes that implements the Fourier matrix, 
\eq 
U_{\text{Fourier},j,k}^{(n)} = \frac{1}{\sqrt n} e^{i \frac{2 \pi}{n} j k} ,
\en
we send three identical particles and one particle with varying degree of distinguishability $x$, as described by the distinguishability matrix
\eq 
\S=\left( \begin{array}{cccc} 
1 & 1 & 1 & x \\
1 & 1 & 1 & x \\
1 & 1 & 1 & x \\
x & x & x & 1 \end{array}
 \right) \label{ourdist} . 
\en 
For example, we can use three horizontally polarized photons and a fourth, vertically polarized one.  The  three indistinguishable (horizontally polarized) photons are prepared in a cyclically symmetric state in the input modes, the partially distinguishable photon is injected into the last mode, 
\eq 
\vec r_{\text{f}} =(1,0,0,1,0,0,1,0,1) .  
\en
For the output event \eq \vec s_{\text{f}}=(0,1,1,0,1,0,0,0,1), \en we combinatorially find  $\mathcal{P}_{\text{dist}}(\vec s_{\text{f}})=4!/9^4  \neq 0$, while applying Eq.~(\ref{bestphysinter}) shows that $\vec s_{\text{f}}$ is fully suppressed for any value of $0 \le x \le 1$, violating (\ref{conjecture}). The prevailing suppression for \emph{all} values of $x$ is surprising:  $x \neq 1$ implies that which-path information is present, since one photon is at least partially distinguishable from the others. 

\begin{figure}[th]
\includegraphics[width=\linewidth,angle=0]{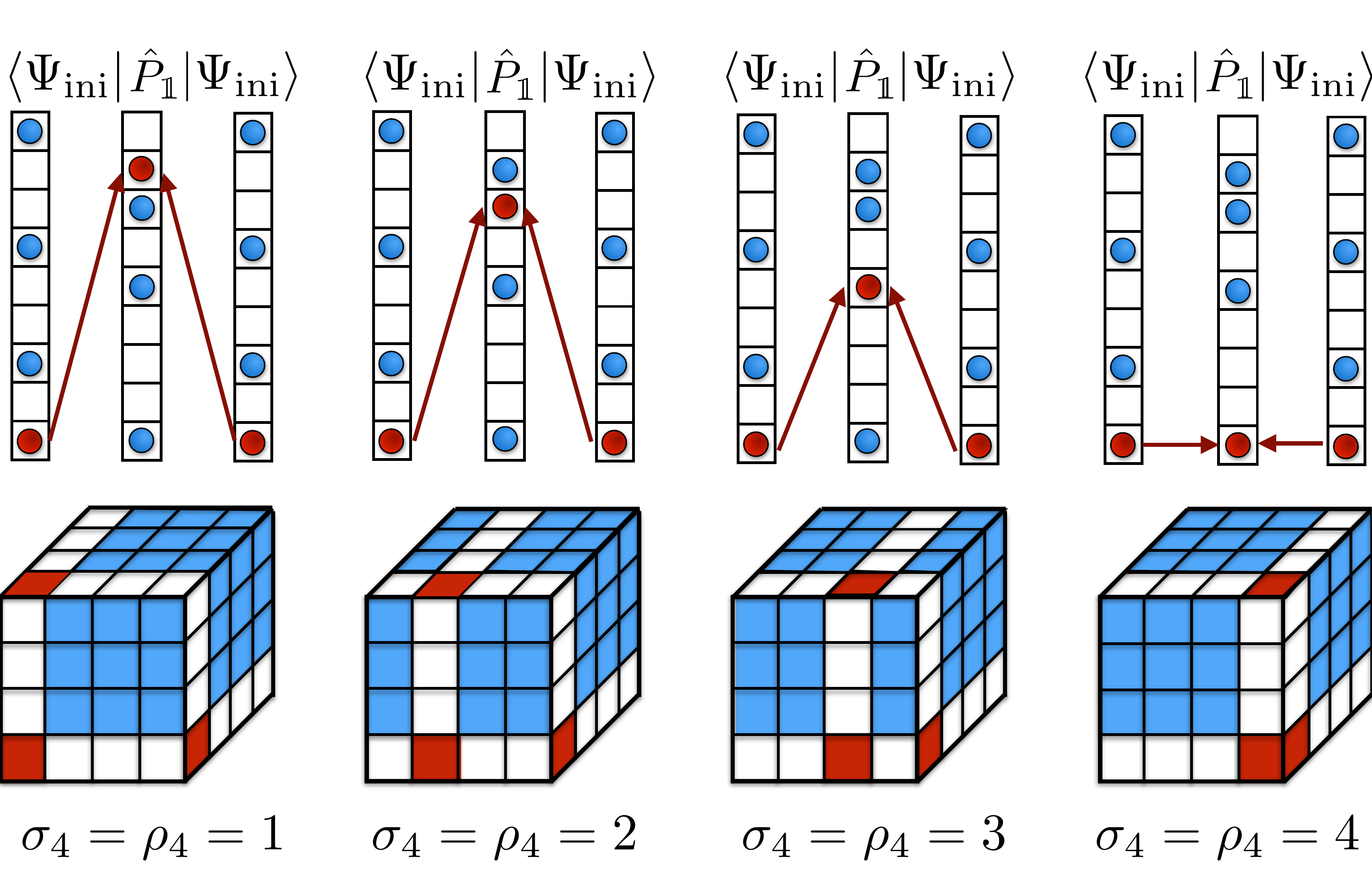}
\caption{(Color online) Double-sided Feynman diagrams and tensor elements contributing to Eq.~(\ref{bestphysinter}),  for the initial state $\vec r_{\text{f}}$ and the final state $\vec s_{\text{f}}$, given the distinguishability matrix (\ref{ourdist}) with $x=0$, i.e.~we visualize the first line of Eq.~(\ref{decompositionSsf}). Four different final states, characterized by the destiny of the red distinguishable particle (mode number $\rho_4=\sigma_4$), are identified, each  is fed by $3!^2$ destructively interfering paths, corresponding to the permutations of the three indistinguishable particles (not shown).  }  \label{FourierSuperSuppression.pdf}
 \end{figure}

Using our formalism, we  understand the suppression as follows: For $x=1$,  the particles are fully indistinguishable and perfectly destructive interference is a consequence of the symmetries of the setup.  For $x=0$, we  apply the suppression law for Fourier matrices \cite{Tichy:2010ZT,Tichy:2012NJP,tichyTutorial} to the three indistinguishable photons. For the arrangement of  indistinguishable bosons $\vec r =(1,0,0,1,0,0,1,0,0)$, the following output configurations are fully suppressed: 
\eq 
 \vec s_1 &=& (0,1,1,0,1,0,0,0,0) , \nonumber  \\
 \vec s_2 &=& (0,0,1,0,1,0,0,0,1) ,  \nonumber \\
  \vec s_3 &=& (0,1,0,0,1,0,0,0,1) ,  \nonumber\\
   \vec s_4 &=& (0,1,1,0,0,0,0,0,1) .    \en 
   Adding a fourth, vertically polarized photon in the last input mode, the four-photon signal $\vec s_{\text{f}}$ also remains suppressed: For any choice of the output mode of the vertically polarized photon (mode numbers 2,3,5,9), the horizontally polarized photons interfere destructively. Formally, in  Eq.~(\ref{bestphysinter}), due to the distinguishability matrix $\S$ defined by Eq.~(\ref{ourdist}), only terms with $\sigma_4=\rho_4$ can be non-vanishing (the particle starting in the fourth mode needs to end in the fourth mode again, since it is distinguishable from all other particles), we therefore obtain four main contributions, visualized in Fig.~\ref{FourierSuperSuppression.pdf}. For each  contribution, the output port of the distinguishable particle is fixed, while the three remaining particles interfere perfectly. Due to symmetry, this interference is always fully destructive, and suppresses the final state $\vec s_{\text{f}}$.

Using Eqs.~(\ref{bestphysinter}),~(\ref{ourdist}), the event probability can be written as
\eq 
\mathcal{P}_{\S}(\vec s_{\text{f}}) &=& \sum_{\substack{ \sigma, \rho \in S_4 \\
 \sigma_4= \rho_4} } \prod_{j=1}^4 M_{\sigma_j ,j} M^*_{\rho_j, j}  \label{decompositionSsf} \\ 
\nonumber &&  + |x|^2 \sum_{ \substack{\sigma, \rho \in S_4 \\ \sigma_4 \neq \rho_4 } } \prod_{j=1}^4 M_{\sigma_j ,j} M^*_{\rho_j, j} , \en
where the first term remains free of exchange contributions that involve the fourth, possibly distinguishable particle. These exchange processes are contained in the second term, which is therefore weighted by the scalar product $|x|^2$. Since the event probability vanishes for $x=0$ and  $x=1$, both sums vanish, and the probability remains 0 for \emph{every} value of $x$. Other examples can be found by a brute-force search that exploits the suppression law \cite{tichyTutorial} to identify candidate instances. A systematic rule to establish such instances remains desirable. 

The above example forces to abandon the intuitive idea that maximum visibility ($V=1$) implies perfect interference: The final event $\vec s$ remains strictly suppressed even though  the interference capability of the system varies. This result harmonizes with the experimental data obtained in Ref.~\cite{Ra:2013kx}, which excludes a naive extrapolation of wave-particle duality to the many-body domain. The question naturally arises whether there exists a scattering setup $U$, a final event $\vec s$ and a distinguishability matrix $\S$ such that $\mathcal{P}_{\S}(\vec s)=0$ while $\mathcal{P}_{\text{id}}(\vec s) \neq 0$, i.e.~whether there can be fully destructive interference exclusively for some configuration of partially distinguishable particles, while a finite probability is associated to fully indistinguishable bosons. Quite counterintuitively, a partially distinguishable many-particle state would then bear stronger interferometric power than a fully indistinguishable one. As shown by the ``zero probability theorem'' of Ref.~\cite{ShchesnovichPartial2014}, however, such setup is impossible.

\subsection{Mixed states} \label{mixedstatessec}
Since the mutual scalar products of different single-mode functions appear directly in the event probability Eq.~(\ref{eq:fullexp}), the latter is straightforwardly extended to mixed states by replacing scalar products by their ensemble average \cite{diffediagnosis}.  Assuming that the particles are uncorrelated, the particle entering the $j$th occupied port populates the state $\ket{\psi_{j,k} }$ with probability $p_{j,k}$, i.e.~it is described by the mixed state
\eq 
\hat \varrho_j=\sum_{k=1}^R p_{j,k} \ket{\psi_{j,k}} \bra{\psi_{j,k}} ,  \label{mixedstates}
\en 
where $R$ denotes the maximal number of states in any pure-state decomposition for all input ports. 
The event probability for mixed states  $\hat \varrho_1, \dots ,\hat  \rho_n$ becomes the ensemble-averaged probability
\eq
\mathcal{P}_{(\hat \varrho_1 \dots \hat  \rho_N )}(\vec s) 
& =& \sum_{k_1 \dots  k_n=1}^R \left(\prod_{j=1}^n p_{j,k_j} \right) \mathcal{P}_{\S[\vec k] } (\vec s) ,   \\
&=&  \left\{  \mathcal{P}_{\S}(\vec s) \right\}_{\hat \varrho_1, \dots, \hat \varrho_n } , 
\en
where $\S[\vec k]$ contains the scalar products associated to the realization $\{k_1, \dots, k_n \}$, occurring with probability $\prod_j p_{j,k_j} $, 
\eq
\S_{j,l}[\vec k]= \braket{\psi_{j,k_j}}{\psi_{l,k_l }}, \en and $\{ \}_{\hat \varrho_1, \dots, \hat \varrho_n}$ denotes the ensemble average over mixed states \cite{diffediagnosis}.  Using Eq.~(\ref{bestphysinter}) and exchanging sums, 
\eq 
\mathcal{P}_{(\hat \varrho_1 \dots \hat \varrho_N )}(\vec s)  =\hspace{5.4cm} \nonumber \\  
\sum_{\sigma, \rho} \sum_{k_1 \dots  k_n} \left(\prod_{j=1}^n \S[\vec k]_{\rho_j, \sigma_j} p_{j,k_j} \right) \prod_{l=1}^n M_{\sigma_l,l} M^*_{\rho_l,l} .
 \en
We set 
 \eq
 J({\sigma, \rho}) &=& \sum_{k_1 \dots  k_n} \left(\prod_{j=1}^n \S[\vec k]_{\rho_j, \sigma_j} p_{j,k_j} \right)  \nonumber \\ 
 &=&  \left\{ \prod_{j=1}^n \S_{\rho_j, \sigma_j}  \right\}_{\hat \varrho_1, \dots, \hat \varrho_n} \label{jmatrix}
 ,
 \en
 for which, for the here assumed ideal detectors, 
 \eq J(\sigma, \rho)&=& J(\rho^{-1}(\sigma), \mathbbm{1}) ,  \label{con1} \\
  J(\sigma, \mathbbm{1}) &=& J^*(\sigma^{-1}, \mathbbm{1}) .   \label{con2} 
  \en
 Using Eq.~(\ref{jmatrix}), we recover the central result of Ref.~\cite{ShchesnovichPartial2014}: 
 \eq 
\mathcal{P}_{(\hat \varrho_1 \dots \hat \varrho_N )}(\vec s) & = &   
 \sum_{\sigma, \rho}  J({\sigma, \rho}) \prod_{l=1}^n M_{\sigma_l,l} M^*_{\rho_l,l} . \label{shchesnovichcentral}
 \en
While the $n \times n$-matrix $\S$ fully describes the interference capability of a pure $n$-particle state, as it pools all $(n-1)n/2$ relevant mutual scalar products, mixed initial states require more physical parameters: 
Each permutation $\sigma$ gives rise to one element of $J$ in Eq.~(\ref{jmatrix}).  Since the ensemble-average of products is not the product of ensemble-averages, e.g., 
 \eq 
 \{ \S_{1,2}  \S_{2,1} \S_{3,4}  \S_{4,3} \}_{\hat \varrho_1, \dots, \varrho_4}  \neq ~~~~~~~ \\ ~~~~~~ \nonumber \{ \S_{1,2}  \S_{2,1} \}_{\hat \varrho_1, \dots, \varrho_4}  \{ \S_{3,4}  \S_{4,3} \}_{\hat \varrho_1, \dots, \varrho_4}  ,
 \en
 the matrix entries of $J$ constitute $n!$ widely independent parameters, constrained by (\ref{con1}), (\ref{con2}). 
  The matrix  $J(\sigma, \rho)$ was introduced in Ref.~\cite{ShchesnovichPartial2014}, including the effect of  non-ideal detectors, which yields further independent physical parameters. For $n=3$ pure photons and ideal detectors, $J$ coincides with the rate matrix (C2) of Ref.~\cite{de-Guise:2014yf}. The rich dependence of event rates on averages of products of scalar products inherent to Eq.~(\ref{shchesnovichcentral}) can be used to diagnose the impact of different decoherence processes on the deterioration of interferometric signals \cite{diffediagnosis}. 

\section{Degree of distinguishability} \label{degreeofdist}

\subsection{Permanent of the distinguishability matrix}
For pure initial states, the representation of the probability as a multi-dimensional permanent (\ref{bigper}) allows us to identify a measure for indistinguishability: The permanent of the distinguishability matrix $\mathcal{S}$. 

The two limiting cases of identical bosons and fully distinguishable particles (Section \ref{distpaidbos}) witness extremal values of $\perm(\S)$. Using  $\mathcal{S}_{j,j}=1$ and $|\mathcal{S}_{j,k}| \le 1$, we find \cite{marcusHadamard,permanents1965}: 
\eq
1 \le \text{perm}(\mathcal{S})  \le n!    \label{inequaperm}
\en
The lower bound constitutes the permanent analogue of the Hadamard determinant inequality, it is only saturated for distinguishable particles ($\mathcal{S}=\mathbbm{1}$); for  multiply occupied input modes, it becomes $ \prod_j r_j!  \le \perm(\S) $. The upper bound is saturated if and only if all particles are fully indistinguishable ($\mathcal{S}=\mathbbm{E}$).  Besides being an unambiguous witness for the extremal cases, $\perm(\S)$ provides a  quantitative indicator for the strength of interference, as shown in the following. 

\subsection{Bounds on the deviation from the idealized cases}
Given a  configuration of internal states $\{ \ket{\phi_1}, \dots , \ket{\phi_n} \}$ associated with  the distinguishability matrix $\S$,  how strongly do the sampling probabilities under $\S$ differ  from those arising for fully distinguishable particles or identical bosons? 
Given an event $\vec s$, if  all mutual scalar products are positive, $\forall j,k: \S_{j,k} \ge 0$, it holds
\eq
|  \mathcal{P}_{\text{dist}}(\vec s) -  \mathcal{P}_{\S}(\vec s) | &\le & \mathcal{P}_{\text{dist}}(\vec s)  \left( \text{perm}(\S) -1 \right) \label{diffromdi} .
\en
For general distinguishability matrices $\S$, 
\eq 
|   \mathcal{P}_{\text{dist}}(\vec s) -  \mathcal{P}_{\S}(\vec s) | &\le & \mathcal{P}_{\text{dist}}(\vec s)  \left( \text{perm}(|\S|) -1 \right) \label{diffromdid} ,
\en 
where the absolute value is taken entrywisely. Under the assumption that $\S$ is real ($\forall j,k: \Im [\S_{j,k} ]= 0$), 
\eq 
|  \mathcal{P}_{\text{id}}(\vec s) -  \mathcal{P}_{\S}(\vec s) |   &\le &  \mathcal{P}_{\text{dist}}(\vec s) \left( n!-\text{perm}(\S)  \right) \label{diffromid} .
\en 

The proofs for these three inequalities are given in Appendix \ref{proofsupperbounds}. By sampling random matrices and states, we found numerical counterexamples against a naive generalization of (\ref{diffromdi})  to general matrices $\S$. No proof for (\ref{diffromid}) for general matrices $\S$ was found, but no numerical counter-example  against such generalization either. Physically,  Eqs.~(\ref{diffromdi}-\ref{diffromid}) quantify the impact of interference contributions on the probability of individual events, i.e.~$\perm(\S)$ quantifies the strength of bosonic exchange contributions in Eq.~(\ref{asnonclass}).

\subsection{Bunching events}
Bunching events of the form $\vec s_{\text{bunch}}=(n, 0 , \dots , 0)$ are particularly strongly affected by indistinguishability. All particles end in the same output port and interfere perfectly constructively, since all columns of the resulting scattering matrix $M$ [Eq.~(\ref{matrixdef})] are identical.  The inequalities (\ref{diffromdi}), (\ref{diffromid}) are then saturated, even for unrestricted distinguishability matrices $\S$: 
 \eq 
\mathcal{P}_{\S}(\vec s_{\text{bunch}}) =  \frac{\text{perm}(\mathcal{S}) }{\prod_{k} r_k!}    \mathcal{P}_{\text{dist}}(\vec s_{\text{bunch}}) , \label{bunching}
\en
where multiple input mode populations are explicitly incorporated to connect our result to the full-bunching law of Refs.~\cite{TichyDiss,PhysRevA.83.062307}, which was formulated for perfectly indistinguishable particles, and experimentally verified in Ref.~\cite{PhysRevLett.111.130503}. In other words, the degree of bunching, i.e.~the factor by which bosonic bunching boosts the probability for a bunching event with respect to  distinguishable particles, is given precisely by the permanent of the distinguishability matrix,  $\perm(\S)$, which thereby becomes a measure for bosonicness. 

\subsection{Total variation distance}
We compare scattering setups in a more holistic way by analyzing the full probability distributions  for all events $\vec s_k$ under a distinguishability matrix $\mathcal{S}$, 
\eq 
\mathcal{\vec  P}_{\mathcal{S}} &=& \left(\mathcal{P}_{\mathcal{S}}(\vec s_0) ,  \mathcal{P}_{\mathcal{S}}(\vec s_1) , \dots \right) , 
\en
where the ordering of events $(\vec s_0, \vec s_1, \dots )$ is irrelevant in our context. 
The total variation distance (or 1-norm) between the probability distributions for identical and partially distinguishable particles is an indicator for their distinctness, 
\eq 
d_{\text{id},\S} \equiv |   \mathcal{\vec P}_{\text{id}} -   \mathcal{\vec P}_{\S} |_1  = \sum_{\vec s}|   \mathcal{\vec P}_{\text{id}}(\vec s) -   \mathcal{\vec P}_{\S}(\vec s)  | , 
\en
which translates in full analogy for the variation distance between the probability distributions for distinguishable and partially distinguishable particles. Naturally, 
\eq 
d_{\mathcal{T}, \S}  \equiv |   \mathcal{\vec P}_{\mathcal{T}} -  \mathcal{\vec P}_{\S} |_1 &\le & 2 , 
\en 
for any two distinguishability matrices $\mathcal{T}$ and $\S$, and the triangle inequality holds, e.g., 
\eq
d_{\text{id},\text{dist}}& \le& d_{\text{id},\S} + d_{\text{dist},\S}. \label{trianglein}
 \en

By taking the sum over all final events $\vec s$ in Eqs.~(\ref{diffromdi},\ref{diffromid}), we can state 
\eq 
\S_{j,k} \ge 0: d_{\text{id},\S} 
   &\le & n!-\text{perm}(\S)  \label{distancefromid} , \\
\Im(\S_{j,k})=0 :  d_{\text{dist},\S} 
 &\le & \text{perm}(\S) -1  \label{distancefromdi} .
\en
We conjecture that  (\ref{distancefromid})  and (\ref{distancefromdi}) remain valid for all distinguishability matrices $\S$, 
but we did not find a proof for these two conjectures; they are motivated by  numerical evidence for random states. 

The bounds can be interpreted as follows: As long as $\perm(\S)$ remains close to unity, interference effects are weak and do not considerably affect the system. The approximation via the classical probability $\mathcal{P}_{\text{dist}}(\vec s)$ is then reliable. On the other hand, $\perm(\S)\approx n!$ is tantamount to almost ideal bosonic interference. For the vast regime between these extremes, however, interference cannot be neglected, but neither can the deviation from the ideal bosonic case. In this realm, the total variation distance to both extremal distributions is large. We numerically found that, on the level of individual event probabilities, the inequalities (\ref{diffromdi},\ref{diffromdid},\ref{diffromid}) are often nearly saturated [in particular, they are saturated for bunching events, Eq.~(\ref{bunching})]; however, when summing over all events $\vec s$, the emerging bounds (\ref{distancefromid},\ref{distancefromdi}) are  inefficient.

\section{Distinguishability transition} \label{visualization}

Being equipped with the analytical tools to study the behavior of many-boson scattering -- in particular, with Eq.~(\ref{ryserbest}) -- we numerically study the behavior of few interfering bosons. Our aim is to dismiss a  simple interpolation between the extreme cases, i.e.~we claim that, in the vast majority of the cases, it is impossible to find a value of $\gamma$ that fulfills \cite{al:2009vn}
\eq \mathcal{\vec P}_{\S} = \mathcal{\vec P}(\gamma) \equiv (1- \gamma) \mathcal{\vec P}_{\text{id}} + \gamma  \mathcal{\vec P}_{\text{dist}} . \label{naiveansatz}
\en
If a  representation of the form (\ref{naiveansatz}) were possible,  partially distinguishable Boson-Sampling would boil down to a simple mixture of classical and Boson-Sampling.  
 To justify our claim, we compute the ``closest'' mixture of the form (\ref{naiveansatz}) for each instance defined by $\S$, i.e.~
\eq
\Delta = \text{min}_\gamma | \mathcal{\vec P}(\gamma)- \mathcal{\vec P}_{\S} |_1 ,
\en
where $\gamma_{\text{best}}$ denotes the optimal value of $\gamma$. The quantity $\Delta$ indicates how ``close'' in distribution space the probability distribution $\mathcal{\vec P}_{\S}$ is to any mixture of $\mathcal{\vec P}_{\text{id}}$ and $\mathcal{\vec P}_{\text{dist}}$, as  sketched in Fig.~\ref{sketchprobsinspace.pdf}(a). 
In other words, ruling out (\ref{naiveansatz}) amounts  to showing that 
 $\Delta>0$ for $1<\perm(\S) < n!$.

\subsection{Canonical transition}
\begin{figure}[h]
\includegraphics[width=\linewidth,angle=0]{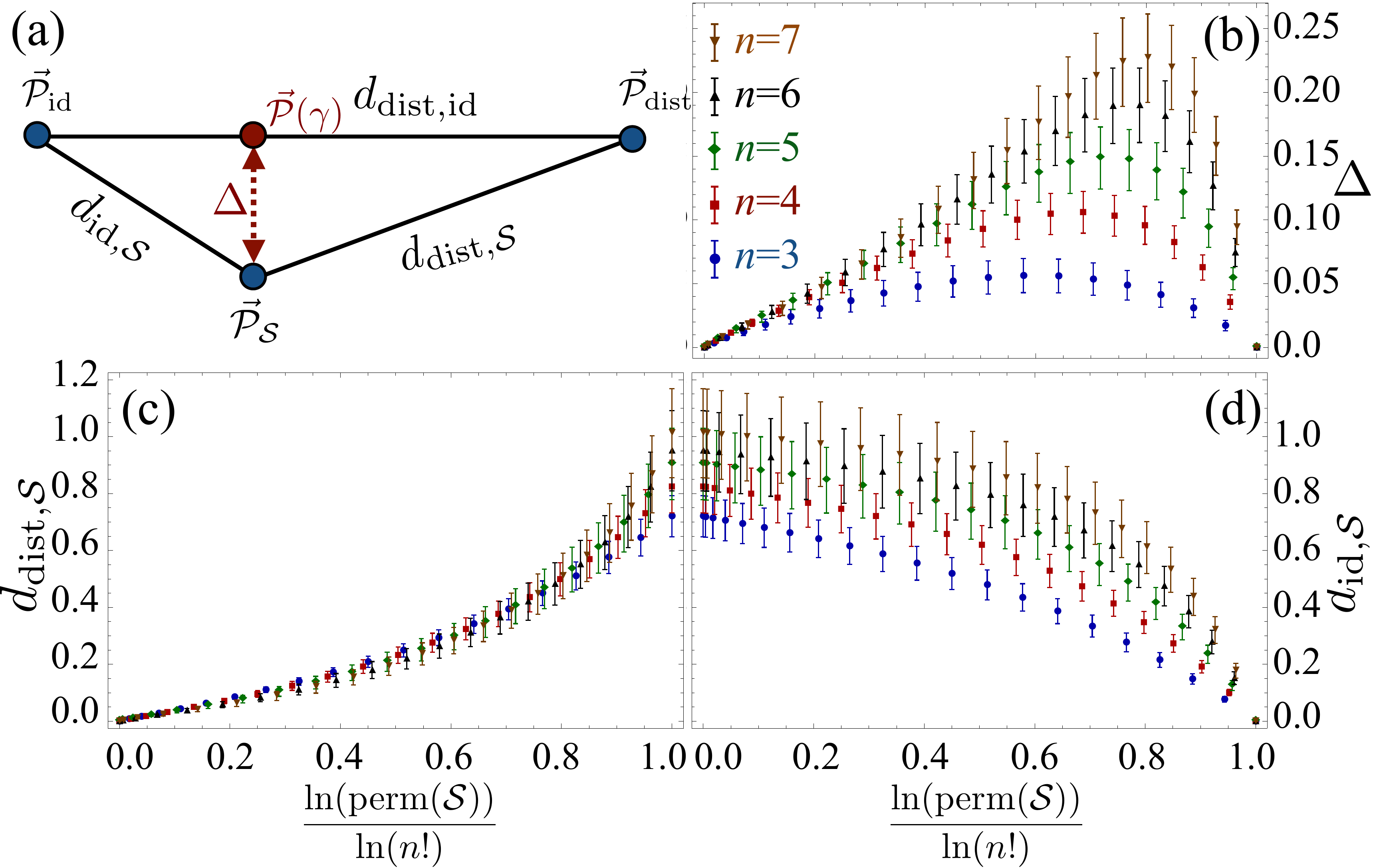}
\caption{(Color online)  (a) Location of $\mathcal{\vec P}_\S$, $\mathcal{\vec  P}_{\text{dist}}$ and $\mathcal{\vec P}_{\text{id}}$ in the high-dimensional space of probability distributions. The value of  $\Delta$ indicates the distance (in 1-norm) to the closest mixture $\mathcal{\vec P}(\gamma)$ on the direct line between $\mathcal{\vec P}_{\text{id}}$ and $\mathcal{\vec P}_{\text{dist}}$.  (b) Minimal total variation distance $\Delta$ between $\mathcal{\vec P}_{\S}$ and a mixture of distinguishable and indistinguishable bosons. (c,d) Total variation distance between $\mathcal{\vec P}_{\S}$ and the probability distribution of distinguishable particles and identical bosons, respectively. We parametrize the many-body state with $n=3,\dots,7$ according to Eq.~(\ref{genericdist}) and show the variation distance as a function of the normalized permanent of $\S$, $0 \le \text{ln}(\perm(\S))/\text{ln}(n!) \le 1$. The error bars show one standard deviation, we have sampled 160 unitary matrices for each $n$ and estimated $\Delta$ by evaluating the probabilities of up to 100 randomly chosen events for each configuration of $x$ and $U$. The number of modes fulfills $m=2n$, 15 equidistant values of $0 \le x \le 1$ were used. $n=3$: blue circles, $n=4$: red squares, $n=5$: green diamonds, $n=5$: black upright triangles and $n=6$: brown downward triangles. } \label{sketchprobsinspace.pdf}
\end{figure}

There are many different ways to proceed from distinguishable particles ($\S=\mathbbm{1}$) to indistinguishable particles ($\S=\mathbbm{E}$) in an experiment \cite{TichyFourPhotons}; we start by studying a smooth parameterization that interpolates between the two limits, 
\eq 
\S_{j,k} = x \text{ for all } j \neq k, \label{genericdist}
\en i.e.~the scalar product between any two single-particle wavefunctions is $x$; 
 the  permanent of $\S$ becomes an $n$th order polynomial in $x$. The matrix $\S$ is realized, e.g., in an $n+1$-dimensional internal Hilbert space in which  the single-particle wavefunctions are given by  
\eq \ket{\phi_k} = (\sqrt{x}, \underbrace{0, \dots, 0}_{k-1}, \sqrt{1-x}, \underbrace{0, \dots, 0}_{n-k-1}) , \en
 in some basis. We explore the transition between bosons and distinguishable particles by varying $x$ in Fig.~\ref{sketchprobsinspace.pdf}(b,c,d). To compare various particle numbers, we show $\Delta$ and the total variation distances $d_{\text{dist},\S}$, $d_{\text{id},\S}$ as a function of the normalized permanent, ${0 \le \text{ln}(\perm(\S))/\text{ln}(n!) \le 1}$. A monotonic relationship between the normalized permanent and the total variation distances emerges: The stronger the interference, the closer we come to the probability distribution for bosons [Fig.~(d)], and detach us from the one for distinguishable particles  [Fig.~(c)]. The former shows a behavior that  depends of the total particle number $n$, for the latter different particle numbers only divide up for large interference, which may, however, be an artifact of our ad-hoc choice of normalization for $\perm(\S)$. The monotonic relationship between the total variation distance and the degree of interference exhibited in Fig.~\ref{sketchprobsinspace.pdf}(c,d) is, however, not realized  on the level of individual events, which often exhibit intricate structures \cite{TichyFourPhotons,Ra:2013kx,ratichycomment,1367-2630-16-1-013006,tichyTutorial}.  Unlike what the data in the two panels (c,d) may suggest at first sight, the probability distribution for partially distinguishable bosons $\mathcal{\vec P}_{\S}$ does not lie on the direct line  between the extremes $\mathcal{\vec P}_{\text{id}}$ and $\mathcal{\vec P}_{\text{id}}$, which is witnessed by the non-vanishing values of $\Delta$ in panel (b). The trend to explore an intricate path in the space of probability distributions far away from any mixture $\mathcal{\vec P}(\gamma)$ becomes more pronounced for larger numbers of particles. 

\subsection{Random states}
To give a more complete picture of the many-body distinguishability transition, we generate instances of $(\gamma_{\text{best}}, \Delta)$ for random choices of $\S$ as follows: For a chosen dimension $2\le D \le n$, the $n$ internal states $\ket{\phi_1}, \dots \ket{\phi_n}$ for the $n$ incoming particles are uniformly randomly chosen. The typical scalar products depend on  $D$: There cannot be $D+1$ fully distinguishable particles in a $D$-dimensional space. 

\begin{figure}[h]
\includegraphics[width=\linewidth,angle=0]{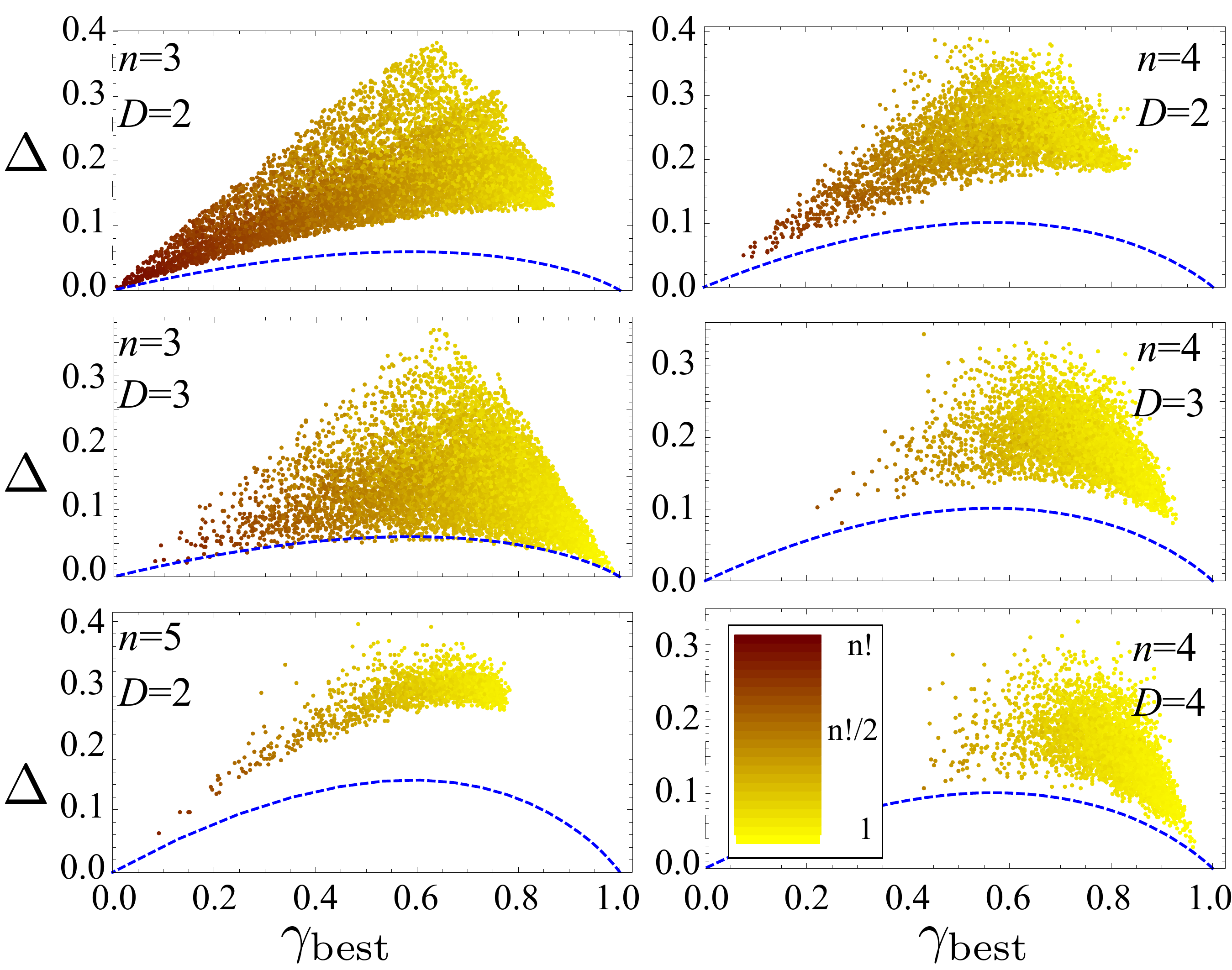}
\caption{(Color online) Randomly sampled instances of  $\gamma_{\text{best}}$ (horizontal axis), $\Delta$ (vertical axis) and $\perm(\S)$ (color). The  internal states of the interfering particles  are randomly chosen in a $D$-dimensional space, for a fixed randomly chosen unitary scattering matrix with $m=2n$. The blue dashed line shows the result for the parametrization  (\ref{genericdist}), where $x$ is varied from 0 (yielding $\gamma_{\text{best}}=1$)  to 1 ($\gamma_{\text{best}}=0$). 
}  \label{visualplaneprobs.pdf}
 \end{figure}

  Our numerical results for randomly chosen unitary matrices of dimensions $m=2n$  are illustrated in Fig.~\ref{visualplaneprobs.pdf}. We sampled $10000$, $5000$ and $2000$ random configurations of distinguishability $\S$ for $n=3,4,5$, respectively, for $D=2, \dots, n$. The color code indicates the degree of interference, $\perm(\S)$. 
Large values of $\perm(\S)$ lead to small values of $\gamma_{\text{best}}$, i.e.~the more the particles interfere, the closer we find ourselves to the ideal situation of indistinguishable bosons. Small values of $\perm(\S)$ are associated to $\gamma_{\text{best}} \approx 1$ and weak interference.  The total variation distance between distinguishable and indistinguishable particles is $d_{\text{id},\text{dist}}=0.70, 0.78, 0.88$ for $n=3,4,5$, respectively, $\Delta$ often takes comparable values. Our parameterization (\ref{genericdist}) typically leads to smaller values of $\Delta$ for a given $\gamma_{\text{best}}$ than for randomly chosen states: The smooth parameterization ensures that all events are equally strongly affected by partial distinguishability, keeping the probability distribution closer to a mixture of the extremal cases.

\subsection{Fourier matrices}
\begin{figure}[h]
\includegraphics[width=.8\linewidth,angle=0]{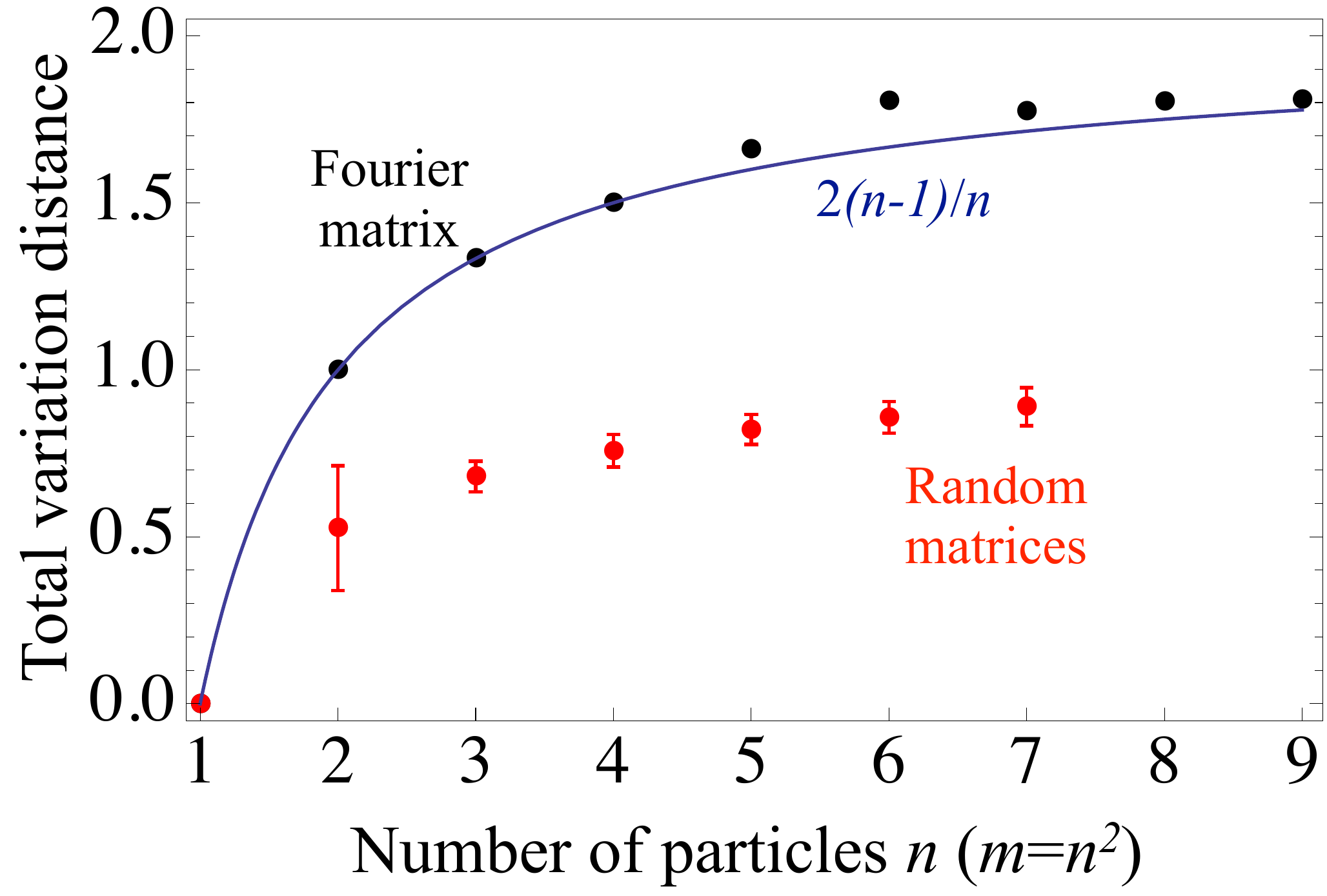}
\caption{(Color online)  Total variation distance between the probability distributions for distinguishable and identical particles. Red points with error bars (one standard deviation): Random matrices, the  average is performed  over 100 random unitary matrices of dimension $m=n^2$, for which the total variation distance is estimated based on a sample of 300 randomly chosen events. Due to computational costs, we only simulated values up to $n=7$. Black points: Fourier matrix with cyclically symmetric initial configuration leading to the suppression of a fraction of approximately $(n-1)/n$  of  events. Blue line: Estimate for the variation distance for the Fourier matrix, $2(n-1)/n$. }  \label{Distances.pdf}
 \end{figure}

\begin{figure}[h]
\includegraphics[width=\linewidth,angle=0]{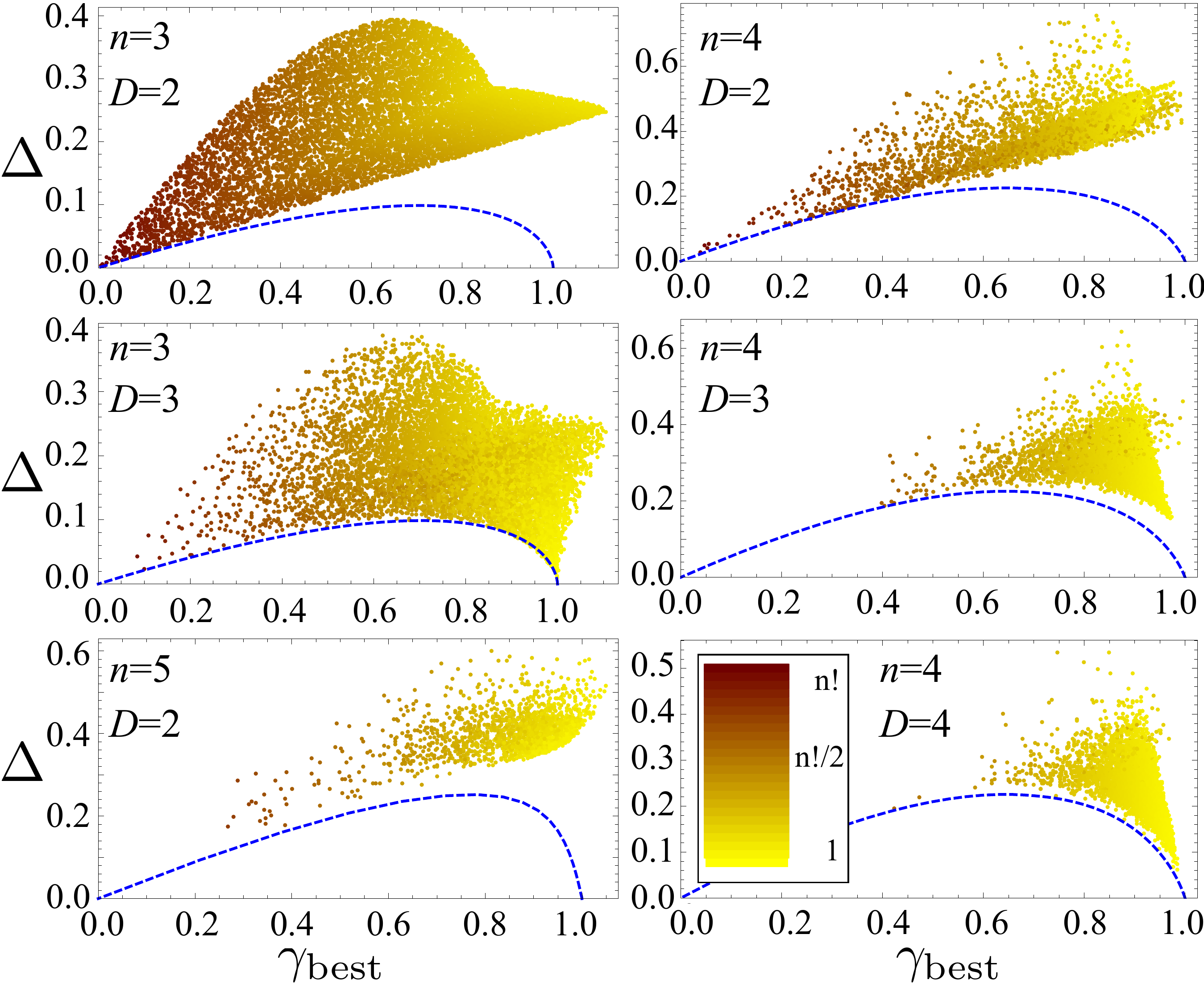}
\caption{(Color online) As in Fig.~\ref{visualplaneprobs.pdf}, but for  Fourier matrices, which come with  larger total variation distances $d_{\text{id},\text{dist}}$ and, consequently, more large values of $\Delta$ compared to random matrices. For some instances of $\gamma_{\text{best}}>1$, the closest distribution $ \mathcal{\vec P}({\gamma})$  contains negative entries and lies outside of the space of probabilities.  }  \label{distancesgeneric.pdf}
 \end{figure}

A particularly sharp difference between distinguishable and identical particles arises for the Fourier-matrix \cite{Tichy:2012NJP}. For a given setting of $n$, $m=2n$ and a cyclically symmetric initial state of the form $\vec r=(1, 0, 0, \dots, 1, 0, 0, \dots)$, the total variation distance between distinguishable and identical particles is significantly larger than for randomly chosen matrices: Due to the Fourier suppression law \cite{Tichy:2010ZT,Tichy:2012NJP}, a fraction of approximately  $(n-1)/n$ of all events is fully suppressed, strongly enhancing the remaining $1/n$ of events. The expected variation distance is, therefore, $2(n-1)/n$, an estimate that  Fig.~\ref{Distances.pdf} confirms empirically. Fourier matrices also lead to particularly large values of $\Delta$ and pronounced edges of the scatter-plot in Fig.~\ref{distancesgeneric.pdf}.  

\subsection{Absence of interpolating  transitions}
Is it ever possible to  parametrize the transition between indistinguishable and distinguishable particles such that we can  write $\mathcal{\vec P}_{\S}$ in the intuitive form (\ref{naiveansatz})? Here, we argue that, beyond a trivial case, such construction is  unlikely. The following mixed many-particles state  leads to (\ref{naiveansatz}):
\eq 
\rho =(1- \gamma ) \ket{0}\bra{0}^{\otimes n} + \gamma \otimes_{j=1}^n \ket{j}\bra{j} , \label{trivialpara}
\en
where $\braket{j}{k} = \delta_{j,k}$ and we refer to the $n+1$ internal states of the particles. Eq.~(\ref{trivialpara}) describes a probabilistic mechanism that produces a state of $n$ indistinguishable particles with probability $\gamma$ and a state of $n$ distinguishable particles otherwise. The particles in this mixed state are strongly classically correlated: They are prepared either all in the  same state or all in different states. 

For uncorrelated pure states, the condition for a parameterization of the form (\ref{naiveansatz}) reads 
\eq
\mathcal{P}_{\S}(\vec s) &=& 
\mathcal{P}_{\text{dist}}(\vec s) + \sum_{\sigma \neq \mathbbm{1}} \left( \prod_{k=1}^n \S_{k,\sigma_k} \right) \perm( M * M_{\sigma,\mathbbm{1}} )  \nonumber  \\
& \stackrel{!}{ =} & \mathcal{P}_{\text{dist}}(\vec s) + (1- \gamma) \sum_{\sigma \neq \mathbbm{1}} \perm( M * M_{\sigma,\mathbbm{1}} ) , \label{sethard}
\en
which has to hold for \emph{all} events $\vec s$ (remember that the matrix $M$ depends on the output state $\vec s$ in Eq.~(\ref{matrixdef})). For a fixed unitary matrix, the number of events $\vec s$ is much larger than the number of adjustable parameters $n(n-1)/2$, such that Eq.~(\ref{sethard}) is heavily overdetermined and a solution impossible in the vast majority of the cases. 

It is difficult to relate the contributions of different permutations $\sigma$ to each other. A neat relationship may be useful to devise a parametrization: Since the sum of all event probabilities is unity, 
\eq 
\sum_{\sigma \neq \mathbbm{1}} \sum_{\vec s} \text{perm}(M * M^*_{\sigma,\mathbbm{1}} ) =0 ,  \label{totalinterferencevanishes}
\en
and the latter  holds for all distinguishability matrices $\S$, it is valid  for each  permutation $\sigma$, such that 
\eq 
\forall \sigma  \neq \mathbbm{1} : \sum_{\vec s} \text{perm}(M * M^*_{\sigma,\mathbbm{1}} ) =0 . \label{suminterferencecontrib}
\en
Intuitively speaking, each exchange process defined by $\sigma\neq \mathbbm{1}$ yields positive and negative interference contributions, which cancel in total when considering all possible events.

 It remains open whether a non-trivial parametrization exists for mixed uncorrelated states of the form (\ref{mixedstates}): Since the ensemble-averaged scalar products take many different independent values (Section \ref{mixedstatessec}), an ingenious parametrization that achieves (\ref{sethard}) is not excluded.

\FloatBarrier

\section{Conclusions and outlook}
The treatment of partially distinguishable particles used implicitly in Ref.~\cite{Shchesnovich2013,Spagnolo:2013fk}, formalized in Ref.~\cite{ShchesnovichPartial2014} and taken further in this article  overcomes the issues enumerated in Section \ref{orthonorsection}: Since our central Eq.~(\ref{bestphysinter}) directly contains the mutual scalar products of the internal states of the interfering particles, there is no dependence on a choice of single-particle basis \cite{tichyTutorial,RohdeNew2014}. Our approach is readily interpreted and visualized by double-sided Feynman diagrams (Fig.~\ref{FullVisualization.pdf}), the emerging multi-dimensional permanent (\ref{bigper}) provides an intuitive and manageable way to deal with the imperfect interference of partially distinguishable bosons.

Simulating Boson-Sampling is classically hard, even if only an approximation is sought, whereas sampling of distinguishable particles can be done efficiently. From this perspective,   the current status is unsatisfactory:    It would be counter-intuitive if the complexity for the \emph{intermediate} case treated here should explode -- as implicit in  Eq.~(\ref{ryserbest}) --while the physical transition between the extremes arises naturally. The immanant-based approach used in Refs.~\cite{Tan:2013ix,de-Guise:2014yf,Tillmann:2014ye} and the smooth behavior of the complexity   of the immanant \cite{buergisser} indeed suggest otherwise. We can nevertheless not conclusively answer our initial question on the computational hardness of the sampling problem in the partially distinguishable realm, but, based on our discussion in Section \ref{visualization}, we  refuted any naive interpolation between the limiting cases and presented counter-intuitive phenomena such as the prevalence of fully destructive interference for partially distinguishable particles (Section \ref{exampleperfectsupp}). 
 
Even though the distinguishability transition is not mediated by a single parameter and can take many different forms depending on the actual path taken from $\S=\mathbbm{1}$ to $\S=\mathbbm{E}$ \cite{TichyFourPhotons}, a measure for interference capability was found: the permanent of the distinguishability matrix $\S$. Not only does it yield the simple bounds (\ref{diffromdi}), (\ref{diffromdid}), and (\ref{diffromid}) on the deviation to the extremal cases, but  it can be read off experimentally as the degree of bunching  [Eq.~(\ref{bunching})]. 
The latter result is connected with the recent observation that Boson-Sampling with thermal states is  related to the permanent of a positive semidefinite hermitian matrix \cite{whatcanqosay}. In our case, the matrix $\S$ defined in Eq.~(\ref{distinguishabilitymatrix}) encodes the mutual distinguishability of the internal states of the injected bosons, while in Ref.~\cite{whatcanqosay}, the pertinent hermitian matrix is built of elements of  the scattering matrix $U$. The permanent of a positive semidefinite hermitian matrix is not a hard computational problem, since one can conceive efficient classical approximations \cite{whatcanqosay}, consistent with the existence of strong bounds \cite{marcusMinc,Minc:1984uq}. 

The role of the permanent is strengthened as an  ubiquitous and essential function in many-boson interference, generalizing the classical and bosonic cases. To compare different experiments and particle numbers, it remains to find an sensible way to scale the range of $\perm(\S)$, $[1,n!]$, such that it quantitatively reflects the degree of interference; one possibility is our normalization adopted in Fig.~\ref{sketchprobsinspace.pdf}. Given that the bounds (\ref{diffromdi}) and (\ref{diffromid}) are formulated under  strong assumptions on the structure of $\S$, it is also desirable to find a generalization  to unrestricted distinguishability matrices $\S$, as well as an extension to mixed states. Although Eq.~(\ref{ryserbest}) is   more benevolent than Eq.~(\ref{orthoproba}), we suspect that there are more efficient ways to evaluate (\ref{bestphysinter}) than our adaptation of Ryser's algorithm: Even in the most general case, the multi-dimensional permanent (\ref{bigper}) is not applied to a general 3-tensor with $n^3$ independent elements, but to a tensor fixed by two complex matrices through Eq.~(\ref{Wtensor}). A way to exploit this symmetry in the computation of probabilities would  significantly alleviate the computational expenses related to partially distinguishable bosons. 

The sampling problem possesses well-understood limiting cases: semi-classical sampling in the many-particle limit $n \gg m$ \cite{Shchesnovich:2013uq,JuanDiego}, classical sampling for distinguishable particles, Fourier-sampling for structured scattering matrices \cite{tichyTutorial,Tichy:2013lq} and computationally hard Boson-Sampling \cite{Aaronson:2011kx}, possibly realized with an input beyond multi-mode Fock-states \cite{Rohde:2013vn,olson2014,sedhardreesan}. These discrete cases constitute the corners of the high-dimensional and widely unexplored \emph{phase diagram of sampling complexity}, whose precise demarkation from a physical and computer-science perspective constitutes an ambitious desideratum. Given the fruitful interplay of computational complexity theory and physics in the understanding of many-particle interference so far, it is not unrealistic to hope for future further synergy and  insight \cite{compstatphys2006}.

\emph{Note added in proof:} Recently, a generalization and improvement of the bounds given in Eqs. (\ref{diffromid}) and (\ref{distancefromid}) were reported in Ref.~\cite{sheshnonew}.

\subsection*{Acknowledgements} The author thanks Scott Aaronson, Clemens Gneiting, Steven Kolthammer, Hyang-Tag Lim, Benjamin Metcalf, Young-Sik Ra, Si-Hui Tan and Ian Walmsley for inspiring discussions, Hubert de Guise, Klaus M\o{}lmer and Valery Shchesnovich for very valuable feedback on the manuscript, and the Villum Foundation and the Danish Council for Independent Research for financial support.

\appendix

\section{Proofs and useful relations}

\subsection{Maximizing permutation} \label{maxiperm}
The most important contribution to the sum (\ref{eq:fullexp}) is the classical event probability: 
\eq
\text{max}_\sigma |\text{perm}(M * M^*_{\sigma,\mathbbm{1}} )| = ~~~~~~~~ \nonumber \\ ~~~~~ \text{perm}(M * M^*_{\mathbbm{1},\mathbbm{1}} ) = \mathcal{P}_{\text{dist}}(\vec s) \label{maxpers}
\en
To see this, set $m_\rho= \prod_{j} M_{j,\rho_j}$ and write
\eq 
\text{perm}(M * M^*_{\sigma, \mathbbm{1}} ) = \sum_{\rho \in S_n} m_\rho m_{\rho(\sigma)}^* ,
\en 
which is maximized for $\sigma=\mathbbm{1}$.

\subsection{Upper bound on probability difference} \label{proofsupperbounds}
We first prove Eq.~(\ref{diffromdi}), where we explicitly require $\S_{j,k} \ge 0$ for all $j,k$: 
\eq 
|  \mathcal{P}_{\text{dist}}(\vec s) &-&   \mathcal{P}_{\S}(\vec s) |  \nonumber \\ & \overset{(\ref{eq:fullexp}), (\ref{distprobfull})}{=}&   \left| \sum_{\sigma \in S_n} \text{perm}(M * M_{\sigma,\mathbbm{1}}^* ) \left( \delta_{\mathbbm{1},\sigma} - \prod_{j=1}^n \S_{j,\sigma_j}  \right) \right|    \nonumber  \\
& =&   \left| \sum_{\sigma \in S_n, \sigma \neq \mathbbm{1}} \text{perm}(M * M_{\sigma,\mathbbm{1}}^* )  \prod_{j=1}^n \S_{j,\sigma_j}   \right|  \nonumber \\
& \le &   \sum_{\sigma \in S_n, \sigma \neq \mathbbm{1}}  \left| \text{perm}(M * M_{\sigma,\mathbbm{1}}^* )  \prod_{j=1}^n \S_{j,\sigma_j}   \right|     \nonumber  \\
& \le &   \text{max}_{\rho} \text{perm}(M * M_{\rho,\mathbbm{1}}^* )  \sum_{\sigma \in S_n, \sigma \neq \mathbbm{1}}   \prod_{j=1}^n \S_{j,\sigma_j}     \nonumber   \\
 & \overset{(\ref{maxpers})}{=}& 
   \mathcal{P}_{\text{dist}}(\vec s) ( \text{perm}(\S)-1)   .
\en
Breaking the assumption $\S_{j,k} \ge 0$  invalidates the inequality. Considering $\perm |\S |$ instead, we recover inequality (\ref{diffromdid}).

Similarly, one  shows (\ref{distancefromid}) for distinguishability matrices fulfilling $\S_{j,k} \in \mathbbm{R}$: 
\eq 
|   \mathcal{P}_{\text{id}}(\vec s) &-&   \mathcal{P}_{\S}(\vec s) |     \nonumber \\ 
&\overset{(\ref{eq:fullexp}), (\ref{idprob})}{=}&  \left| \sum_{\sigma \in S_n} \text{perm}(M * M_{\sigma,\mathbbm{1}}^* ) \left( 1  - \prod_{j=1}^n \S_{j,\sigma_j}  \right) \right|   \nonumber \\
&\le &   \text{max}_{\rho} \text{perm}(M * M_{\rho,\mathbbm{1}}^* )  \sum_{\sigma \in S_n} | 1- \prod_{j} \S_{j,\sigma_j} |  \nonumber \\
& \overset{(\ref{maxpers})}{=}&  \mathcal{P}_{\text{dist}}(\vec s) \left(n! - \text{perm}(\S) \right)  \label{disteqqid}
\en
For general matrices $\S$, we have not found any instance that violates the generalization of inequality (\ref{disteqqid}), but the last step of our proof above  breaks down.


\end{document}